\documentclass[11pt]{JHEP3}
\usepackage{subfigure,axodraw, amsmath,graphicx}

\title{Generating MHV super-vertices in light-cone gauge}

\author{ Chih-Hao Fu\\
Department of Mathematical Sciences, University of Durham\\
South Road, Durham, DH1 3LE, U.K.\\ 
E-mails:
 \email{chih-hao.fu@durham.ac.uk}} 

\preprint{DCPT-09/85}

\abstract{ 
We constructe the $\mathcal{N}=1$ SYM lagrangian in light-cone gauge using chiral superfields instead of the standard vector superfield approach 
and derive the MHV lagrangian. The canonical transformations of the gauge field and gaugino fields are summarised by the transformation condition of  chiral superfields. We show that  $\mathcal{N}=1$ MHV super-vertices can be described by a formula similar to that of the $\mathcal{N}=4$ MHV super-amplitude. In the discussions we briefly remark on how to derive Nair's formula for  $\mathcal{N}=4$ SYM theory directly from  light-cone lagrangian.
}

\begin{document}
\section{Introduction}

Ever since its discovery the CSW prescription originally conjectured
by Cachazo, Svr{\v c}eck and Witten in \cite{CSW,twistor string}
has been shown to be an efficient method for constructing gluon scattering
amplitudes. In this approach the off-shell continued MHV amplitudes
(non-trivial amplitudes with the largest number of positive helicity gluons) are
taken as the new vertices, which allows numerous varieties of subgraph
structures in the standard Feynman graphs to be represented by Parke-Taylor
formula \cite{Parke Taylor}. The CSW rules were successfully generalised
to one-loop level and to include quarks and superpartners \cite{CSW loop,CSW SUSY,CSW fermion,Bern:2009xq}.
A lagrangian derivation of the rules was found by Mansfield \cite{MHV lagrangian} and independently by Gorsky and Rosly \cite{Gorsky:2005sf} 
through canonically transforming the self-dual part of the LCYM lagrangian
into a free field theory

\begin{equation}
\mathcal{L}^{-+}\left[\mathcal{A}\right]+\mathcal{L}^{-++}\left[\mathcal{A}\right]=\mathcal{L}^{-+}\left[\mathcal{B}\right].\end{equation}

The transverse components $\mathcal{A}$ and $\bar{\mathcal{A}}$
of the gauge field in light-cone coordinates were assumed to be functionals
of the new field variables $\mathcal{B}$ and $\bar{\mathcal{B}}$
so that after performing the transformation the vertices in the new
lagrangian have the same helicity structure as prescribed by the CSW
rules. 

\begin{eqnarray}
&& \mathcal{A}_{1}=\mathcal{B}_{1}+\Upsilon_{123}\mathcal{B}_{2}\mathcal{B}_{3}+\cdots\label{eq:A}  \\
&& 
\hat{\partial}\bar{\mathcal{A}}_{1}=\hat{\partial}\bar{\mathcal{B}}_{1}+\Xi_{123}^{2}\hat{\partial}\bar{\mathcal{B}}_{2}\mathcal{B}_{3}+\Xi_{123}^{3}\mathcal{B}_{2}\hat{\partial}\bar{\mathcal{B}}_{3}+\cdots\label{eq:A bar}
\end{eqnarray}

In 4-dimensions the MHV vertices were algebraically verified to agree
with the Parke-Taylor formula using holomorphy of the translation
kernels $\Upsilon$ and $\Xi^{k}$ in \cite{Ettle Morris, n point MHV vertices}
and the D-dimensional theory was given in \cite{us 1} which restored
the loop-level amplitudes originally appeared ``missing'' from the
CSW rules. The corresponding MHV lagrangians for QCD and SQCD were
derived by Ettle, Morris and Xiao by tranforming the physical field
components and their canonical conjugate variables on a pair by pair
basis \cite{James MHV QCD,Xiao SQCD}.

An alternative strategy was found by Britto, Chachazo, Feng and Witten
from analysing singularities of the amplitude when external leg momenta
are shifted by a complex value \cite{BCF,BCFW}. Using Cauchy's theorem
it was shown that a generic amplitude can be derived from scattering
amplitudes of fewer particles whose leg momenta are determined by
poles. The method of BCFW recursion has been shown to be a powerful
tool for tree-level calculations \cite{BCFW tree} and was extended
to theories containing massive particles and fermions \cite{BCFW massive,BCFW fermion}.
Combining with generalised unitarity the BCFW recursion relation was
also extended to loop-level calculations \cite{BCFW unitarity}. 

Recently, BCFW recursion has been generalised to compute
tree-level amplitudes in $\mathcal{N}=4$ supersymmetry Yang-Mills
theory \cite{BCFW SUSY}. Instead of shifting individual scattering
ampiltudes labeled by particle species and momenta, in the supersymmetry
generalisation of BCFW recursion one considers super-amplitudes
whose initial and final states are described by momentum space super-wavefunctions.
The super-wavefunction contains a superposition of all the single
particle states in the $\mathcal{N}=4$ supermultiplet, each of them
being tagged by bookkeeping Grassmann variables $\eta_{A}$,

\begin{eqnarray}
&& \Phi(p,\eta)=G^{+}(p)+\eta_{A}\Gamma^{A}(p)+\frac{1}{2}\eta_{A}\eta_{B}S^{AB}(p)+\frac{1}{3!}\eta_{A}\eta_{B}\eta_{C}\epsilon^{ABCD}\bar{\Gamma}_{D}(p)
 \nonumber \\
&& \hspace{2cm} +\frac{1}{4!}\eta_{A}\eta_{B}\eta_{C}\eta_{D}\epsilon^{ABCD}G^{-}(p).\label{super wvfunct}
\end{eqnarray}

The conventional scattering amplitude is obtained from differentiating
 super-amplitude with respect to the appropriate Grassmann variables
determined by the species of the particles which participate the scattering
event. For example differentiating Nair's formula for  $\mathcal{N}=4$
MHV super-amplitude \cite{Nair} 

\begin{equation}
A_{N=4}^{MHV}(1,2,\cdots n)=\frac{1}{\left\langle 12\right\rangle \left\langle 23\right\rangle \cdots\left\langle n1\right\rangle }\delta^{(4)}(\sum_{i,j=1}^{n}\left\langle i\, j\right\rangle \eta_{Ai}\eta_{Aj})\label{Nair N4 formula 2}\end{equation}

with respect to $\eta_{Ai}$ and $\eta_{Aj}$, from $A=1$ to $4$,
yields the familiar Parke-Taylor formula for gluon scattering, where
the $i$-th and $j$-th legs are associated with negative helicity
gluons. The generalised supersymmetry BCFW recursion formula is then
obtained by applying Cauchy's theorem to the super-amplitude where
Grassmann variables are shifted along with leg momenta \cite{BCFW SUSY travaglini}.
All tree-level super-amplitudes were computed in \cite{All tree}
by Drummond and Henn and the super-amplitudes were verified to be
superconformal invariant, where the SUSY generators for $\mathcal{N}=4$
theory were given by \cite{twistor string},

\begin{equation}
Q_{\alpha\, A}=\lambda_{\alpha}\eta_{A},\hspace{0.4cm}\bar{Q}_{\dot{\alpha}}^{A}=\bar{\lambda}_{\dot{\alpha}}\frac{\partial}{\partial\eta_{A}}.\label{N4 SUSY gen 1}\end{equation}

In this paper we canonically transform the chiral superfields in the usual momentum space description to produce
the MHV lagrangian for $\mathcal{N}=1$ SYM theory. The supersymmetry
generalisations to equations (\ref{eq:A}) and (\ref{eq:A bar}) automatically
summarise the relation between the transformation formulae of 
gluon and gluino fields. We perform a fermionic integral transformation which
 replaces the super-space Grassmann variables $\theta$ and $\bar{\theta}$ used to label 
 off-shell superfields by a single Grassmann variable $\eta$. The integral transformation
allows us to directly derive the $\mathcal{N}=1$ analougue of (\ref{N4 SUSY gen 1})
as representations of the SUSY generators in the new super-space.
In section \ref{N1 BCFW} we adapt the supersymmetry BCFW recursion
to  $\mathcal{N}=1$ theory and calculate the generic n-point ``helicity-ordered'' MHV
super-amplitude formula. The more symmetrical formula 
derived by Bern, Carrasco, Ita, Johansson and Roiban \cite{Bern:2009xq} can be
reproduced by superimposing helicity-ordered super-amplitudes with all possible  
helicity configurations. We compute the translation kernels to all
order and show that the MHV super-vertices in the new lagrangian are
described by the same formula as that of the MHV super-amplitude

\begin{equation}
V_{N=1}^{MHV}(1^{+},2^{+}\dots i^{-},j^{-},\dots n^{+})=\frac{\left\langle i\, j\right\rangle ^{3}}{\left\langle 12\right\rangle \left\langle 23\right\rangle \cdots\left\langle n1\right\rangle }\delta(\sum_{i,j=1}^{n}\left\langle i\, j\right\rangle \eta_{i}\eta_{j}),\label{N1 Parke Taylor}\end{equation}

with the off-shell continuation defined in the light-cone $\check{p}$ direction 
which is absent in formula (\ref{N1 Parke Taylor}).  
In the discussions we remark on how the method is extended to the $N=4$ SYM
theory to generate the super-space MHV lagrangian that has vertices described by Nair's
formula (\ref{Nair N4 formula 2}). The notation used throughout this paper and,
in particular, the definition of spinors for off-shell momenta
are summarised in appendix \ref{appendix notation}.

\section{Chiral Construction of the $\mathcal{N}=1$ SYM lagrangian}

\label{section chiral N1}

In the textbook approach, the supersymmetric non-abelian gauge theory
is constructed from the fieldstrength of a vector superfield. To arrive
at an MHV lagrangian we can pick a convenient gauge, integrate over
unphysical degrees of freedom, and canonically transform the helicity
gluon and gluino fields seperately. However the underlying supersymmetry
implies that we can organise the physical fields that belong to the
same supermultiplet into a conceptually simpler structure. In \cite{Feng:2006yy}
 Feng and Huang successfully derived the MHV lagrangian for $\mathcal{N}=4$ 
SYM theory in which the chiral superfield has a simple transformation condition 
in coordinate space. 
In persuit of this idea we may be tempted to apply the canonical transformation
directly on the $\mathcal{N}=1$ vector superfield, but the vector superfield depends
on a large number of unphysical degrees of feedom and the lagrangian
does not have an easily manipulated structure when it is written in
terms of vector superfields. As noted by Siegel and Gates \cite{Siegel:1981ec}
alternatively the physical field components can be packed into chiral superfields. 
A systematic method was developed by Ananth, Brink,
Lingren, Nilsson and Ramond to derive the $\mathcal{N}=4$ light-cone
SYM lagrangian dimensionally reduced from 10-dimensions \cite{Brink,Brink Ramond}.
In light-cone gauge the physical fields are
closed under the SUSY subalgebra $Q_{1}$, $\bar{Q}_{\dot{1}}$.
\begin{eqnarray}
&&
Q_{1}\mathcal{A}=i\Lambda,\hspace{0.4cm} \bar{Q}_{\dot{1}}\mathcal{A}=0,\hspace{0.4cm} Q_{1}\bar{\mathcal{A}}=0,\hspace{0.8cm}\bar{Q}_{\dot{1}}\bar{\mathcal{A}}=-i\bar{\Lambda},\label{LC SUSY transform 1} \\
&&
Q_{1}\Lambda=0,\hspace{0.4cm}\bar{Q}_{\dot{1}}\Lambda=\hat{\partial}\mathcal{A},\hspace{0.4cm} Q_{1}\bar{\Lambda}=-\hat{\partial}\bar{\mathcal{A}},\hspace{0.4cm}\bar{Q}_{\dot{1}}\bar{\Lambda}=0,\label{LC SUSY transform 2}
\end{eqnarray}

where in the above equations we used $\psi_{\alpha}=\left(\begin{array}{c}
\bar{T}\\
\bar{\Lambda}\end{array}\right)$, $\bar{\psi}_{\dot{\alpha}}=\left(\begin{array}{c}
T\\
\Lambda\end{array}\right)$ to denote the gluino field components. Note that with the quadratic
terms eliminated by the gauge condition $\hat{\mathcal{A}}=0$, the
transformation relations (\ref{LC SUSY transform 1}) and (\ref{LC SUSY transform 2})
remain the same in the on-shell process. The closure of the physical field components under this SUSY subalgebra
allows us to define chiral superfields without the need of introducing
auxiliary fields  \cite{Brink}.
\begin{eqnarray}
&&
\Phi(x,\theta)=\mathcal{A}(y)+i\theta\,\Lambda(y),\label{chiral phi} \\
&&
\bar{\Phi}(x,\theta)=\bar{\mathcal{A}}(\bar{y})+i\bar{\theta}\,\bar{\Lambda}(\bar{y}),\label{chiral phi bar}
\end{eqnarray}

where gluons and gluinos having the same helicities are enclosed into
the same supefield,  $y=(x^{+},\, x^{-}+\frac{1}{2}i\theta\bar{\theta},\, x^{z},\, x^{\bar{z}})$, and
 $\bar{y}=(x^{+},\, x^{-}-\frac{1}{2}i\theta\bar{\theta},\, x^{z},\, x^{\bar{z}})$.
We introduce the shorthand notation for representations of the SUSY
covariant derivatives and generators $D_{1}=d$, $\bar{D}_{\dot{1}}=\bar{d}$,
$Q_{1}=q$, and $\bar{Q}_{\dot{1}}=\bar{q}$, which stand for

\begin{equation}
d=\frac{\partial}{\partial\theta}+\frac{i}{2}\bar{\theta}\hat{\partial}\,,\,\bar{d}=-\,\frac{\partial}{\partial\bar{\theta}}-\frac{i}{2}\,\theta\hat{\partial},\label{SUSY cov de}\end{equation}

\begin{equation}
q=\frac{\partial}{\partial\theta}-\frac{i}{2}\bar{\theta}\hat{\partial}\,,\bar{\, q}=-\,\frac{\partial}{\partial\bar{\theta}}+\frac{i}{2}\,\theta\hat{\partial}.\label{SUSY gen}\end{equation}

The superfields defined in (\ref{chiral phi}) and (\ref{chiral phi bar})
satisfy the chiral constraints $\bar{d}\,\Phi=d\,\bar{\Phi}=0$.

Using the chiral superfields just defined, one can construct a SUSY
invariant lagrangian as a D-term integral \cite{Belitsky:2004sc}

\begin{equation}
S=\frac{-4i}{g^{2}}\int d^{4}x\, d\theta\, d\bar{\theta}\, tr\left(\bar{\Phi}\frac{\partial^{2}}{\hat{\partial}}\Phi+[\Phi,\frac{\bar{\partial}}{\hat{\partial}}\Phi]\bar{\Phi}+[\bar{\Phi},\frac{\partial}{\hat{\partial}}\bar{\Phi}]\Phi-i\left[\Phi,\bar{d}\,\bar{\Phi}\right]\frac{1}{\hat{\partial}^{2}}\left[\bar{\Phi},d\,\Phi\right]\right)\label{N1 SYM lagrangian}\end{equation}

It is straightforward to verify that after integrating $\theta$,
$\bar{\theta}$ equation (\ref{N1 SYM lagrangian}) is the same as
the standard $\mathcal{N}=1$ SYM lagrangian with $\hat{\mathcal{A}}$
eliminated by light-cone gauge condition and all of the unphysical
fields integrated out.  

\section{Transforming light-cone gauge SYM into the new representation}

\label{Grassmann Fourier transform}

 In order to interpret coefficients in the interaction terms as
vertices in Feynman rules we need to work in an unconstrained
description so that the superfields can be taken as functional
integral variables. Note that when $(y,\,\theta,\,\bar{\theta})$
and $(\bar{y},\,\theta,\,\bar{\theta})$ are regarded as sets of independent
variables the chiral conditions demand that the chiral and anti-chiral
superfield to be independent of $\bar{\theta}$ and $\theta$ respectively:
$\bar{d}\Phi=\left.\frac{\partial}{\partial\bar{\theta}}\Phi\right|_{y,\,\theta}=0$,
$d\bar{\Phi}=\left.\frac{\partial}{\partial\theta}\Phi\right|_{\bar{y},\,\bar{\theta}}=0$.
As pointed out by Mandelstam in \cite{Mandelstam:1982cb} for this purpose one
can start with a single Grassmann variable representation and reconstruct
the light-cone gauge action. Nevertheless in this paper we choose
to remove the constraint by mapping the expressions (\ref{chiral phi})  and (\ref{chiral phi bar})
through integral transformations. This is possible because when considered
individually each superfield has only one Grassmannian dependence. It
is the appearance of both chiral and anti-chiral superfields in the
action that requires both $\bar{\theta}$ and $\theta$. The mapping
generalises to terms that contain multiple superfields and to the
MHV lagrangian to be discussed in section \ref{canonical transform superfield}. 
For functions of $(\bar{y},\,\bar{\theta})$,
we define a super-space analogue to the Fourier transform and maps
the function to a new super-space labeled by momentum and a Grassmann
variable $\eta$. Denoting the inverse of this transform as $T$ we
have $T^{-1}:\, f(\bar{y},\,\bar{\theta})\rightarrow f(p,\,\eta)$.
Since SUSY covariant derivatives anti-commute with themselves, the
function $\bar{d}\, f(\bar{y},\,\bar{\theta})$ satisfies anti-chiral
constraint and therefore is a function of $(y,\,\theta)$. This allows
us to reuse the integral transformation to bring both functions $f(\bar{y},\,\bar{\theta})$
and $f(y,\,\theta)$ into the same super-space. $(\bar{d}\, T)^{-1}:\, f(y,\,\theta)\rightarrow f(p,\,\eta)$.
The new super-space $(p,\,\eta)$ has one Grassmann variable fewer
than the original super-space $(x,\,\theta,\,\bar{\theta})$ so the
superfields are no longer constrained after the transformation. Applying
different transformations on chiral and anti-chiral superfields does
not generate extra representations to the SUSY generators here. For
anti-chiral superfields the generator becomes $T^{-1}q\, T$, while
for chiral superfields we have 

\begin{equation}
(\bar{d}\, T)^{-1}q\,(\bar{d}\, T)=-(\bar{d}\, T)^{-1}\bar{d}\, q\, T=-(\bar{d}\, T)^{-1}(\bar{d}\, T)\, T^{-1}q\, T,\end{equation}

and the extra minus sign cancels after moving generator in the new
representation across the Grassmannian delta function produced
 from $(\bar{d}\, T)^{-1}(\bar{d}\, T)$.

We choose the integral transformation for an anti-chiral superfield
as

\begin{equation}
\bar{\Phi}(\bar{y},\,\bar{\theta})=-\int d^{4}\bar{y}\, d\eta\, e^{-ip\cdot\bar{y}}\delta(\bar{\theta}\hat{p}^{\frac{1}{2}}-\eta)\,\bar{\phi}(p,\,\eta)\label{eq:yspace}\end{equation}

The transformation for chiral superfields follows from taking covariant
derivative on both sides of (\ref{eq:yspace}). In momentum space
the formulae read:%
\footnote{When an analytic continuation for the negative light-cone energy $\hat{p}$ is needed, we
use
\begin{equation}
\bar{\Phi}(p,\,\theta)=-\int d\eta\, e^{-\frac{1}{2}\theta\bar{\theta}\hat{p}}\delta(\bar{\theta}\sqrt{\hat{p}}-\eta)\,\bar{\phi}(p,\,\eta),\label{sqrt phi bar}\end{equation}

\begin{equation}
\bar{d}\,\bar{\Phi}(p,\,\theta)=-\sqrt{\hat{p}}\int d\eta\, e^{\frac{1}{2}\theta\bar{\theta}\hat{p}}\delta(1-\theta\eta\hat{p}/\sqrt{\hat{p}})\,\bar{\phi}(p,\,\eta)\label{sqrt d phi bar}\end{equation}

in place of (\ref{transform phi bar}) and (\ref{transform d phi bar}),
 where the square roots are determined by the analytic continuation 
used in the definition of Lorentz invariant bracket $\langle ij \rangle =(ij)/\sqrt{\hat{i}}\sqrt{\hat{j}}$.
In the rest of this paper we follow the convention used in \cite{Bern:2009xq} and
define $\sqrt{\hat{p}} = sgn(\hat{p}) \left|\hat{p}\right|^{\frac{1}{2}}$. Alternatively one can 
adopt the analytic continuation of \cite{Gunion:1985vca} and assume $\sqrt{\hat{p}}=i\left|\hat{p}\right|^{\frac{1}{2}}$ when $\hat{p}$ is negative, 
which produces a relative phase factor $i$
in the delta function in propagator (\ref{free super L}).%
}

\begin{equation}
\bar{\Phi}(p,\,\theta)=-\int d\eta\, e^{-\frac{1}{2}\theta\bar{\theta}\hat{p}}\delta(\bar{\theta}\hat{p}^{\frac{1}{2}}-\eta)\,\bar{\phi}(p,\,\eta),\label{transform phi bar}\end{equation}

\begin{equation}
\bar{d}\,\bar{\Phi}(p,\,\theta)=-\hat{p}^{\frac{1}{2}}\int d\eta\, e^{\frac{1}{2}\theta\bar{\theta}\hat{p}}\delta(1-\theta\eta\hat{p}^{\frac{1}{2}})\,\bar{\phi}(p,\,\eta),\label{transform d phi bar}\end{equation}

\begin{equation}
\Phi(p,\,\theta)=\int d\eta\, e^{\frac{1}{2}\theta\bar{\theta}\hat{p}}\delta(1-\theta\eta\hat{p}^{\frac{1}{2}})\,\phi(p,\,\eta),\label{transform phi}\end{equation}

\begin{equation}
d\,\Phi(p,\,\theta)=-\hat{p}^{\frac{1}{2}}\int d\eta\, e^{-\frac{1}{2}\theta\bar{\theta}\hat{p}}\delta(\bar{\theta}\hat{p}^{\frac{1}{2}}-\eta)\,\phi(p,\,\eta)\label{transform d phi}\end{equation}

The inverse of the integral transformations above are given by

\begin{equation}
\phi(p,\,\eta)=-\hat{p}^{-\frac{1}{2}}\int d\theta e^{-\frac{1}{2}\theta\bar{\theta}\hat{p}}\delta(1+\theta\eta\hat{p}^{\frac{1}{2}})\,\Phi(p,\,\theta),\label{inverse phi}\end{equation}

\begin{equation}
\bar{\phi}(p,\,\eta)=-\frac{1}{\hat{p}}\int d\theta e^{-\frac{1}{2}\theta\bar{\theta}\hat{p}}\delta(1+\theta\eta\hat{p}^{\frac{1}{2}})\,\bar{d}\,\bar{\Phi}(p,\,\theta),\label{inverse phi bar}\end{equation}

and the momentum space superfields in the new representation are
\begin{eqnarray}
 &&
\phi(p,\eta)=i\bar{\Lambda}(p)\hat{p}^{\frac{1}{2}}+\eta\mathcal{A}(p),\label{phi} \\
&&
\bar{\phi}(p,\eta)=\bar{\mathcal{A}}(p)+\eta i\Lambda(p)\hat{p}^{-\frac{1}{2}}.\label{phi bar}
\end{eqnarray}

In the same way that the symmetry generators of the Poincare group
can be converted to momentum space, the representation of the SUSY
generators in the light-cone coordinates (\ref{SUSY gen}) can be
expressed in terms of $\eta$.

\begin{equation}
q=\hat{p}^{\frac{1}{2}}\eta,\hspace{0.4cm} \bar{q}=\hat{p}^{\frac{1}{2}}\frac{\partial}{\partial\eta} \label{eta rep}\end{equation}

It is straightforward to verify that after neglecting
all of the commutators appearing in the SUSY transformation relations,
for asymptotic states we have

\begin{equation}
Q_{\alpha}=\lambda_{\alpha}\eta,\hspace{0.4cm}  \bar{Q}_{\dot{\alpha}}=\bar{\lambda}_{\dot{\alpha}}\frac{\partial}{\partial\eta},\label{onshell SUSY gen}\end{equation}

which are the $\mathcal{N}=1$ versions of the on-shell SUSY generators
introduced by Witten in \cite{twistor string}. Note however, that the
 representation (\ref{eta rep}) holds even when off-shell.

\subsection{3-point MHV and $\overline{\text{MHV}}$ vertices}

The light-cone gauge SYM lagrangian can be quickly rewritten in terms
of the new representation by applying (\ref{transform phi bar}) to
(\ref{transform d phi}) to its field contents. In addition to a momentum
conservation delta function, the free field part contains a Grassmanian
 delta function, which forbids interchanges between gluons and gluinos 
in the absence of an interaction.

\begin{equation}
S_{free}=\int d^{4}x\, d\theta d\bar{\theta}\,\mathcal{L}^{-+}=\int d^{4}p_{1}\, d^{4}p_{2}\, d\theta\, d\bar{\theta}\,\bar{\Phi}\frac{p_{2}^{2}}{\hat{p}_{2}}\delta^{(4)}(p_{1}+p_{2}) \nonumber \Phi\end{equation}

\begin{equation}
=\int d^{4}p_{1}\, d^{4}p_{2}\, d\eta_{1}d\eta_{2}\bar{\phi}(p_{1},\eta_{1})\delta^{4}(p_{1}+p_{2})\delta(\eta_{1}+\eta_{2})\, p_{2}^{2}\,\phi(p_{2},\eta_{2})\label{free super L}\end{equation}

For the $(--+)$ interaction term we apply the same transformation
again to have

\begin{equation}
\int d^{4}x\, d\theta d\bar{\theta}\,\mathcal{L}^{--+}=tr\int d^{4}x\, d\theta d\bar{\theta}[\bar{\Phi},\frac{\partial}{\hat{\partial}}\bar{\Phi}]\Phi \hspace{3cm} \nonumber \end{equation}

\begin{equation}
=tr\int d^{4}p_{1}\dots d\eta_{1}\dots\frac{\left(12\right)}{\hat{1}\hat{2}}\,\hat{3}^{\frac{1}{2}}(\hat{3}^{\frac{1}{2}}\eta_{1}\eta_{2}+\hat{1}^{\frac{1}{2}}\eta_{2}\eta_{3}+\hat{2}^{\frac{1}{2}}\eta_{3}\eta_{1})\,\bar{\phi}_{1}\bar{\phi}_{2}\phi_{3}\label{3pt N1 MHV}\end{equation}

Using momentum conservation condition we can combine all of the hat-components
and the round bracket into spinor brackets. Note that the vertex factor
resembles the super-amplitude formula given by Nair for $\mathcal{N}=4$
SYM theory.

\begin{equation}
tr\int d^{4}p_{1}\dots d\eta_{1}\dots\frac{\left\langle 12\right\rangle ^{3}}{\left\langle 12\right\rangle \left\langle 23\right\rangle \left\langle 31\right\rangle }\left(\left\langle 12\right\rangle \eta_{1}\eta_{2}+\left\langle 23\right\rangle \eta_{2}\eta_{3}+\left\langle 31\right\rangle \eta_{3}\eta_{1}\right)\,\bar{\phi}_{1}\bar{\phi}_{2}\phi_{3}\label{3pt N1 MHV 2}\end{equation}

We then compute the $(++-)$ term to give

\begin{equation}
\int d^{4}x\, d\theta d\bar{\theta}\,\mathcal{L}^{++-}=tr\int d^{4}x\, d\theta d\bar{\theta}[\Phi,\frac{\bar{\partial}}{\hat{\partial}}\Phi]\bar{\Phi} \hspace{4cm} \nonumber \end{equation}

\begin{equation}
=tr\int d^{4}p_{1}\dots d\eta_{1}\dots\frac{\left\{ 12\right\} }{\hat{1}\hat{2}}\,\hat{3}^{\frac{1}{2}}(\hat{1}^{\frac{1}{2}}\eta_{1}+\hat{2}^{\frac{1}{2}}\eta_{2}+\hat{3}^{\frac{1}{2}}\eta_{3})\,\phi_{1}\phi_{2}\bar{\phi}_{3}\label{3pt N1 MHV bar} \hspace{2cm} \nonumber \end{equation}

\begin{equation}
=tr\int d^{4}p_{1}\dots d\eta_{1}\dots\frac{\left[12\right]^{3}}{\left[12\right]\left[23\right]\left[31\right]}\left(\left[23\right]\eta_{1}+\left[31\right]\eta_{2}+\left[12\right]\eta_{3}\right)\,\phi_{1}\phi_{2}\bar{\phi}_{3}\label{3pt N1 MHV bar 2}\end{equation}

In the defining equations for integral transformations (\ref{transform phi bar})
to (\ref{transform d phi}) we chose to substitue the $\bar{\theta}\hat{p}^{\frac{1}{2}}$
dependence in $\bar{\Phi}$ and $d\,\Phi$ by the newly introduced
Grassmann variable $\eta$. As an alternative we can choose our definition
to substitue the $\theta\hat{p}^{\frac{1}{2}}$
in $\Phi$ and $\bar{d}\,\bar{\Phi}$. In that case we will have the
same spinor bracket factors for the 3-point MHV vertex (\ref{3pt N1 MHV 2}) but with
the sum $\left(\left\langle 12\right\rangle \eta_{1}\eta_{2}+\left\langle 23\right\rangle \eta_{2}\eta_{3}+\left\langle 31\right\rangle \eta_{3}\eta_{1}\right)$ replaced by
$\left(\left\langle 12\right\rangle \eta_{3}+\left\langle 23\right\rangle \eta_{1}+\left\langle 31\right\rangle \eta_{2}\right)$, and similarly for the $\overline{\text{MHV}}$ term. 
The same correspondence between the $\eta$ assignment of
the super-wavefunction (\ref{super wvfunct}) and the $\eta$ dependence
in the super-amplitudes for $\mathcal{N}=4$ SYM was found by
Drummond, Henn, Korchemsky and Sokatchev in \cite{All tree,N4 SUSY Henn Korchemsky 1,N4 SUSY Henn Korchemsky 2}.

The remaining two 4-point vertices can be calculated following the
same procedure.

\begin{equation}
V_{1234}^{2}=\frac{1}{\left(\hat{2}+\hat{3}\right)^{2}}\left(\left(\hat{1}\hat{3}+\hat{2}\hat{4}\right)\eta_{1}\eta_{2}-\hat{1}^{\frac{1}{2}}\hat{3}^{\frac{1}{2}}\left(\hat{2}+\hat{3}\right)\eta_{2}\eta_{3}+\hat{2}^{\frac{1}{2}}\hat{3}^{\frac{1}{2}}\left(\hat{1}+\hat{4}\right)\eta_{1}\eta_{3}\right. \nonumber \end{equation}
\begin{equation}
\hspace{3cm} \left. +2\,\hat{1}^{\frac{1}{2}}\hat{2}^{\frac{1}{2}}\hat{3}^{\frac{1}{2}}\hat{4}^{\frac{1}{2}}\eta_{3}\eta_{4}-\hat{2}^{\frac{1}{2}}\hat{4}^{\frac{1}{2}}\left(\hat{1}+\hat{4}\right)\eta_{4}\eta_{1}+\hat{1}^{\frac{1}{2}}\hat{4}^{\frac{1}{2}}\left(\hat{2}+\hat{3}\right)\eta_{2}\eta_{4}\right),\label{4pt N1 1}\end{equation}

and 

\begin{equation}
V_{1234}^{3}=-\left(\frac{\hat{3}\hat{4}}{\left(\hat{1}+\hat{4}\right)^{2}}+\frac{\hat{1}\hat{4}}{\left(\hat{3}+\hat{4}\right)^{2}}\right)\eta_{1}\eta_{3}-\frac{\hat{1}^{\frac{1}{2}}\hat{2}^{\frac{1}{2}}\hat{4}}{\left(\hat{3}+\hat{4}\right)^{2}}\eta_{2}\eta_{3}-\frac{\hat{2}^{\frac{1}{2}}\hat{3}^{\frac{1}{2}}\hat{4}}{\left(\hat{1}+\hat{4}\right)^{2}}\eta_{1}\eta_{2} \nonumber \end{equation}
\begin{equation}
+\hat{3}^{\frac{1}{2}}\hat{4}^{\frac{1}{2}}\left(\frac{1}{\left(\hat{1}+\hat{4}\right)}-\frac{\hat{1}}{\left(\hat{3}+\hat{4}\right)^{2}}\right)\eta_{4}\eta_{1}+\hat{4}^{\frac{1}{2}}\hat{1}^{\frac{1}{2}}\left(\frac{1}{\left(\hat{3}+\hat{4}\right)}-\frac{\hat{3}}{\left(\hat{1}+\hat{4}\right)^{2}}\right)\eta_{3}\eta_{4} \nonumber \end{equation}
\begin{equation}
 -\hat{1}^{\frac{1}{2}}\hat{2}^{\frac{1}{2}}\hat{3}^{\frac{1}{2}}\hat{4}^{\frac{1}{2}}\left(\frac{1}{\left(\hat{1}+\hat{4}\right)^{2}}+\frac{1}{\left(\hat{3}+\hat{4}\right)^{2}}\right)\eta_{2}\eta_{4},\hspace{2cm} \label{4pt N1 2}\end{equation}

where we used $V_{1234}^{2}$ and $V_{1234}^{3}$ to denote the vertices
that have adjacent and next-to-adjacent negative helicity legs.

\begin{equation}
\int d^{4}x\, d\theta d\bar{\theta}\,\mathcal{L}^{--++}=\frac{4}{g^{2}}\int d^{4}x\, d\theta d\bar{\theta}\left[\Phi,\bar{d}\,\bar{\Phi}\right]\frac{1}{\left(i\hat{\partial}\right)^{2}}\left[\bar{\Phi},d\,\Phi\right] \nonumber \end{equation}
\begin{equation}
=\frac{4}{g^{2}}\, tr\int V_{1234}^{2}\bar{\phi}_{1}\bar{\phi}_{2}\phi_{3}\phi_{4}+V_{1234}^{3}\bar{\phi}_{1}\phi_{2}\bar{\phi}_{3}\phi_{4} \hspace{3cm} \end{equation}

\section{Calculating super-amplitudes using functional methods}

\label{section super Feynman rules}

One of the advantages of rewriting  light-cone $\mathcal{N}=1$
SYM lagrangian in terms of chiral superfields is that it allows us
to compute super-amplitudes from a set of manifestly supersymmetric
Feynman rules. In a system where supersymmetry is effectively unbroken
it is reasonable that particles in the same same supermultiplet can
be treated as a single entity. Nevertheless, in the standard functional
integral approach to Green function calculations there is a fundamental
distinction between a field and the field of its superpartner. The
gluon fields are bosonic while the gluino fields are taken as fermionic,
both fields are regarded as independent variables to be integrated
over in the functional integral. A naive attempt to combine these
two fields by a change of variables reduces the degrees of freedom
and does not make sense mathematically. So instead of on integration
variables we focus on the generating functional that generates Green
functions.

\begin{equation}
Z\left[J\right]=\int\mathcal{DA}\mathcal{D}\bar{\mathcal{A}}\mathcal{D}\Lambda\mathcal{D}\bar{\Lambda}\, e^{iS+i\int\, j^{(A)}\mathcal{A}+\bar{\mathcal{A}}j^{(\bar{A})}+j^{(\Lambda)}\Lambda+\bar{\Lambda}j^{(\bar{\Lambda})}},\end{equation}

where $S$ is the $\mathcal{N}=1$ SYM action (\ref{N1 SYM lagrangian})
in light-cone gauge, and we introduce generating currents $j^{(A)}$,
$j^{(\bar{A})}$, $j^{(\Lambda)}$ and $j^{(\bar{\Lambda})}$ for
every physical fields. We note that the current term integral can
be simplify by the introduction of super-currents

\begin{equation}
\int d^{4}p\, j^{(A)}\mathcal{A}+\bar{\mathcal{A}}j^{(\bar{A})}+j^{(\Lambda)}\Lambda+\bar{\Lambda}j^{(\bar{\Lambda})}=\int d^{4}p\, d\eta\, J\phi+\bar{\phi}\bar{J},\end{equation}

where we defined

\begin{equation}
J=j^{(A)}-i\hat{p}^{\frac{1}{2}}\eta j^{(\Lambda)},\hspace{0.4cm} \bar{J}=\eta j^{(\bar{A})}-i\hat{p}^{\frac{1}{2}}j^{(\bar{\Lambda})}.\label{super current}\end{equation}

As in the standard calculation we extract the interaction terms as
variation operators of the generating currents. From equations (\ref{3pt N1 MHV})
to (\ref{4pt N1 2}) we saw the interactions in the lagrangian can
be written as functionals of the momentum space superfields. This
means that using the chain rules the variations with respect to the
currents associated with physical fields can be combined as $-i\frac{\delta}{\delta J}$
and $i\frac{\delta}{\delta\bar{J}}$, and the vertices are simply
given by equations (\ref{3pt N1 MHV}) to (\ref{4pt N1 2}), where
the variation with respect to a function of both bosonic and fermionic
variables is defined as in \cite{super Feynman rules 1} \begin{equation}
\frac{\delta J(p^{'},\eta^{'})}{\delta J(p,\eta)}=\delta^{4}(p^{'}-p)\,\delta(\eta^{'}-\eta)\end{equation}

Therefore we have

\begin{equation}
Z\left[J\right]=e^{iS_{int}\left[\frac{\delta}{\delta J}\right]}Z_{0}\left[J\right]\label{interaction functional}\end{equation}

The free generating functional is calculated from integrating over
all field variables.

\begin{equation}
Z_{0}\left[J\right]=e^{i\int\, j^{(\bar{A})}\Delta_{(A)}j^{(A)}+j^{(\bar{\Lambda})}\Delta_{(\Lambda)}j^{(\Lambda)}}=e^{i\int\,\bar{J}\Delta J}\label{free gen functional}\end{equation}

where the light-cone gauge gluon and gluino propagators are given
by $\Delta_{(A)}(p_{1},\, p_{2})=\frac{1}{p_{2}^{2}}\delta^{4}(p_{1}+p_{2})$,
$\Delta_{(\Lambda)}(p_{1},\, p_{2})=\frac{\hat{p}_{2}}{p_{2}^{2}}\delta^{4}(p_{1}+p_{2})$.
We find that the currents associated with gluons and gluinos can be
again organised into the super-currents (\ref{super current}), and
the propagators of two different particle species are replaced by

\begin{equation}
\Delta(p_{1},\eta_{1};\, p_{2},\eta_{2})=\frac{1}{p_{2}^{2}}\delta^{4}(p_{1}+p_{2})\,\delta(\eta_{1}+\eta_{2})\label{super prop}\end{equation}

Note that despite the free generating functional (\ref{free gen functional})
was derived without treating the chiral superfields as field variables,
the propagator (\ref{super prop}) takes the form as the inverse of
the free superfield lagrangian (\ref{free super L}), allowing us
to introduce superfields $\phi$ and $\bar{\phi}$ as auxiliary field
variables, where we generalised the functional integral to fields
labeled by both bosonic and fermionic indices $p$ and $\eta$ in
the same way as in \cite{super Feynman rules 1,super Feynman rules 2}
so that the integration over $\phi$ and $\bar{\phi}$ has the same
properties as over ordinary fields. The interaction part of the action
extracted as a functional of variation operators can be applied back
to the $Z_{0}\left[J\right]$ to restore the lagrangian as a functional
of $\phi$ and $\bar{\phi}$. It is easy to see that the lagrangian
has the same propagator and vertices given in (\ref{free super L})
to (\ref{4pt N1 2}).

\begin{equation}
Z_{0}\left[J\right]=\int\mathcal{D}\phi\mathcal{D}\bar{\phi}\, e^{iS_{free}+\int\, J\phi+\bar{\phi}\bar{J}},\, Z\left[J\right]=\int\mathcal{D}\phi\mathcal{D}\bar{\phi}\, e^{iS+i\int\,\bar{J}\Delta J}\label{superfield gen functional}\end{equation}

For the purpose of computing the generating functional and the Green
function it makes no different whether the functional integral was
defined from the physical fields or from the superfield viewpoint.

Using the standard Wick contrction and the LSZ reduction on (\ref{interaction functional})
and (\ref{free gen functional}), it is straightforward to derive
a set of supersymmetric Feynman rules for $\mathcal{N}=1$ SYM theory
based on the super-momentum space lagrangian, and the method naturally
leads to a combination of scattering amplitudes related to each other
by supersymmetry transformation. Therefore  we define the ``helicity-ordered'' 
super-amplitude as the LSZ reduction of the superfield Green function

\begin{equation}
A(p_{i},\sigma_{i},\eta_{i})=\lim_{p^{2}_{i}\rightarrow}\prod_{i}p_{i}^{2}\left\langle \cdots\phi\cdots 
\bar{\phi}
\cdots
\right\rangle, 
\label{superamplitude}\end{equation}

at the expense of manifest CPT symmetry comparing with the definition used in \cite{Bern:2009xq}. 
To convert the super-amplitude into the physical scattering amplitudes
of gluons and gluinos we extract terms with the Grassmann variables
corresponding to the particle species participating the event. From
the definitions of superfields (\ref{phi}), (\ref{phi bar}) we see
a Grassmannian momentum $\eta_{i}$ is present whenever there is a
positive helicity gluon or a negative helicity gluino. The appropriate
polarisations factors for the LSZ reduction formula are automatically
included from the definition of a super-amplitude (\ref{superamplitude}).

\begin{equation}
\left\langle 1^{+}\cdots2_{\Lambda}^{+}\cdots3^{-}\cdots4_{\Lambda}^{-}\right\rangle 
 =\frac{\partial}{\partial\eta_{4}}\cdots \frac{\partial}{\partial\eta_{1}}
\prod_{i}p_{i}^{2}\langle \mathcal{A}_{1}\cdots\frac{\Lambda_{2}}{\hat{p}_{2}^{\frac{1}{2}}}\dots\bar{\mathcal{A}}_{3}\cdots\frac{\bar{\Lambda}_{4}}{\hat{p}_{4}^{\frac{1}{2}}}\rangle 
\label{superamplitude 2}\end{equation}

Note that the superfields $\phi$ and $\bar{\phi}$ here are regarded
as auxiliary fields introduced in the functional integral (\ref{superfield gen functional})
which do not contain physical gluon or gluino fields as components.
The expansion from the definition of a super-amplitude (\ref{superamplitude})
into a series of physical scattering amplitudes 
relies on current algebra. However since the chain rule of variations
does not distinguish whether the current $j^{(A)}-i\hat{p}^{\frac{1}{2}}\eta j^{(\Lambda)}$
is multiplied by the combination $i\bar{\Lambda}(p)\hat{p}^{\frac{1}{2}}+\eta\mathcal{A}(p)$
or the newly introduced integration variable $\phi(p,\eta)$, the
scattering amplitude calculated from integrating over gluon and gluino
fields is the same as the amplitude calculated from integrating over
$\phi(p,\eta)$.

\subsection{Applying BCFW to calculate $\mathcal{N}=1$ MHV super-amplitudes}

\label{N1 BCFW}

In \cite{BCFW SUSY,BCFW SUSY travaglini,N4 SUSY Henn Korchemsky 1,N4 SUSY Henn Korchemsky 2}
the BCFW recursion method is generalised to $\mathcal{N}=4$ SYM theory
to compute super-amplitudes that have super-wavefunctions as end states.
We adapt the argument provided by Brandhuber, Heslop and Travaglini
originally designed to apply on super-amplitudes having two positive
helicity gluon lines shifted in the $\mathcal{N}=4$ theory \cite{BCFW SUSY travaglini}
to super-amplitudes with positive and one negative leg shifted in
the $\mathcal{N}=1$ theory and derive the formula for 4-point MHV
super-amplitude.

\begin{figure}[!h]
\centering
     \subfigure{
%
  \begin{picture}(150,78) (24,-9)
    \SetWidth{0.375}
    \Line(50,30)(25,5) 
    \Line(50,30)(25,60)  

    \Line(50,30)(130,30)
    \Line(130,30)(155,5)  
    \Line(130,30)(155,60)  
    \SetWidth{0.5}

    \BCirc(50,30){14}
    \GCirc(130,30){14}{0.75}  
                              \Text(62,27)[br]{$MHV$}
                              \Text(145,27)[br]{$\overline{\text{MHV}}$}  
    
        \Text(25,0)[br]{$-$} 
        \Text(25,-10)[br]{$2$} 
        
         \Text(25,60)[br]{$-$} 
         \Text(30,70)[br]{$1^{'}$} 
        
        \Text(165,0)[br]{$+$} 
        \Text(170,-10)[br]{$3$} 
          \Text(165,65)[br]{$+$}  
          \Text(160,75)[br]{$4^{'}$}  
                \Text(75,35)[br]{$+$}  
                \Text(95,40)[br]{$q^{'}$}  
                \Text(115,35)[br]{$-$} 

                
  \end{picture} 
    
  }   \caption{Shifting the helicity-ordered super-amplitude $A(1^{-},2^{-},3^{+},4^{+})$}
  \label{Fig SUSY BCFW}
\end{figure}
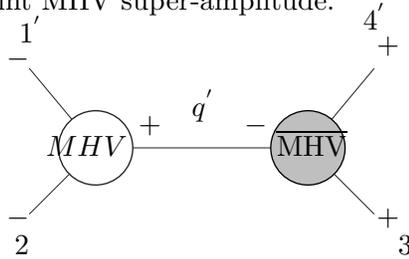

Consider shifting the leg 1 and 4 of the super-amplitude $A(1^{-},2^{-},3^{+},4^{+})$
(Fig.\ref{Fig SUSY BCFW}). The momenta $p_{1}$ and $p_{4}$ are shifted in
the same way as in the pure Yang-Mills theory,
\begin{eqnarray}
&&
P_{1\,\alpha\dot{\alpha}}^{'}(z)=\lambda_{1\,\alpha}\bar{\lambda}_{1\,\dot{\alpha}}-z\,\lambda_{1\,\alpha}\bar{\lambda}_{4\,\dot{\alpha}}, \nonumber \\ 
&& P_{4\,\alpha\dot{\alpha}}^{'}(z)=\lambda_{4\,\alpha}\bar{\lambda}_{4\,\dot{\alpha}}+z\,\lambda_{1\,\alpha}\bar{\lambda}_{4\,\dot{\alpha}}.\label{SUSY shift 1}
\end{eqnarray}

In addition to momenta we also shift the Grassmann variable associated
with the negative helicity leg as

\begin{equation}
\eta_{1}^{'}(z)=\eta_{1}-z\,\eta_{4},\label{SUSY shift 2}\end{equation}

while all other momenta and Grassmann variables are unchanged. A super-amplitude
defined in (\ref{superamplitude}) generically contains a series of
physical amplitudes, each of them being multiplied by the corresponding
Grassmann variables. The ratios between scattering amplitudes of different
particle species are fixed by the SUSY Ward identities. Because the
shiftings given by equations (\ref{SUSY shift 1}) and (\ref{SUSY shift 2})
leave the SUSY generators invariant,

\begin{equation}
\bar{Q}_{\dot{\alpha}}^{'}=\sum_{i=1}^{4}\bar{\lambda}_{i\,\dot{\alpha}}^{'}\frac{\partial}{\partial\eta_{i}^{'}}=\bar{Q}_{\dot{\alpha}},
\hspace{0.4cm}
 Q_{\alpha}^{'}=\sum_{i=1}^{4}\lambda_{\alpha}^{'}\eta_{i}^{'}=Q_{\alpha},\end{equation}

these ratios can be shown to be independent of the complex variable
$z$. The physical amplitudes contained in the super-amplitude therefore
have the same $z$ dependence as the pure gluon scattering amplitude
and the super-amplitude vanishes asymptotically as $z\rightarrow\infty$.
For example for the MHV super-amplitude $A(1^{-},2^{-},3^{+},4^{+})$,
we solve the ratios explicitly by repeatedly applying SUSY Ward identity
with different SUSY transformation parameters, and the super-amplitude
is proportional to

\begin{equation}
\left\langle 12\right\rangle \eta_{1}\eta_{2}+\left\langle 23\right\rangle \eta_{2}\eta_{3}+\left\langle 34\right\rangle \eta_{3}\eta_{4}+\left\langle 41\right\rangle \eta_{4}\eta_{1}+\left\langle 13\right\rangle \eta_{1}\eta_{3}+\left\langle 24\right\rangle \eta_{2}\eta_{4},\end{equation}

which is invariant under the shifting (\ref{SUSY shift 1}) and (\ref{SUSY shift 2}).
From the above argument we also see that the generalisation to $\mathcal{N}=1$
SYM does not introduce new singluarities, therefore we have, from
the BCFW recursion,

\begin{equation}
A_{4}(0)=\int d\eta_{q}\, d\eta_{q^{'}}\, \left. A_{L}(z)\,\frac{\delta(\eta_{q}+\eta_{q^{'}})}{q^{2}}\, A_{R}(z) \right|_{z=-\left\langle 34\right\rangle /\left\langle 31\right\rangle }\label{SUSY BCFW} \hspace{2cm} \end{equation}

\begin{equation}
=\frac{\left\langle 12\right\rangle ^{2}}{\left\langle 2\, q^{'}\right\rangle \left\langle q^{'}\,1\right\rangle }\,\frac{1}{\left\langle 34\right\rangle \left[34\right]}\,\frac{\left[34\right]^{2}}{\left[4\, q^{'}\right]\left[q^{'}\,3\right]} \hspace{4cm} \nonumber \end{equation}

\begin{equation}
\times\left(\begin{array}{c}
\langle 12\rangle [34]\eta_{1}^{'}\eta_{2}-\langle 2\, q^{'}\rangle [4\, q^{'}]\eta_{2}\eta_{3}-\langle 2\, q^{'}\rangle [q^{'}\,3]\eta_{2}\eta_{4}\\
+\langle q^{'}\,1\rangle [4\, q^{'}]\eta_{1}^{'}\eta_{3}+\langle q^{'}\,1\rangle [q^{'}\,3]\eta_{1}\eta_{4}\end{array}\right),\end{equation}

where $q=p_{3}+p_{4}$. Simplifying the above expression gives the
4-point super-amplitude

\begin{equation}
A(1^{-},2^{-},3^{+},4^{+})=\frac{\left\langle 12\right\rangle ^{3}}{\left\langle 12\right\rangle \left\langle 23\right\rangle \left\langle 34\right\rangle \left\langle 41\right\rangle }\left(\sum_{i,j=1}^{4}\left\langle i\, j\right\rangle \eta_{i}\eta_{4}\right).\end{equation}

Since the argument for asymptotic behavior and the algebraic derivation
we used do not depend on the number of legs, we can replace the amplitude
$A_{L}(z)$ on the left hand side of the propagator by an $(n-1)$-point
MHV super-amplitude. By induction an n-point MHV super-amplitude is
given by the formula (\ref{N1 Parke Taylor}). The BCFW recursion
also extends beyond MHV super-amplitudes because the argument for
asmptotic behaviour only rely on the fact that the SUSY generators
are invariant under the shifting (\ref{SUSY shift 1}) and (\ref{SUSY shift 2}). 

\section{Super-space canonical transformation}

\label{canonical transform superfield}

So far we have derived a supersymmetry equivalent to the LCYM theory.
A natural next step is to perform a canonical transformation on the
field variables as in the pure Yang-Mills theory \cite{MHV lagrangian}
to absorb the unwanted $\overline{\text{MHV}}$ term so that in terms of
the new variables the lagrangian automatically generates CSW rules
for $\mathcal{N}=1$ SYM theory. In \cite{Xiao SQCD} Morris and
Xiao applied the canonical transformation pair by pair. In the gauge
field sector the gluon, gluino fields and their canonical conjugate
momenta $\left\{ \mathcal{A},\hat{\partial}\bar{\mathcal{A}}\right\} $
and $\left\{ \Lambda,\bar{\Lambda}\right\} $ were transformed into
the corresponding new fields $\left\{ \mathcal{B},\hat{\partial}\bar{\mathcal{B}}\right\} $
and $\left\{ \Pi,\bar{\Pi}\right\} $ according to the following expansions 
\begin{eqnarray}
 &&
\mathcal{A}_{1}=\mathcal{B}_{1}+\Upsilon_{123}\mathcal{B}_{2}\mathcal{B}_{3}+\cdots,\label{SQCD 1} \\
&& 
\hat{\partial}\bar{\mathcal{A}}_{1}=\hat{\partial}\bar{\mathcal{B}}_{1}+\Xi_{123}^{2}\hat{\partial}\bar{\mathcal{B}}_{2}\mathcal{B}_{3}+\Xi_{123}^{3}\mathcal{B}_{2}\hat{\partial}\bar{\mathcal{B}}_{3}+\cdots\\
&&
\hspace{1cm} +\Xi_{123}^{2}\bar{\Pi}_{2}\Pi_{3}+\Xi_{123}^{3}\Pi_{2}\bar{\Pi}_{3}+\Xi_{1234}^{2}\bar{\Pi}_{2}\Pi_{3}\mathcal{B}_{4}+\Xi_{1234}^{2}\bar{\Pi}_{2}\mathcal{B}_{3}\Pi_{4}+\cdots, \label{SQCD 2} \\
&&
\Lambda_{1}=\Pi_{1}+\Upsilon_{123}\Pi_{2}\mathcal{B}_{3}+\Upsilon_{123}\mathcal{B}_{2}\Pi_{3}+\cdots,\label{SQCD 3} \\
&&
\bar{\Lambda}_{1}=\bar{\Pi}_{1}+\Xi_{123}^{2}\bar{\Pi}_{2}\mathcal{B}_{3}+\Xi_{123}^{3}\mathcal{B}_{2}\bar{\Pi}_{3}+\cdots.\label{SQCD 4}
\end{eqnarray}

In order to keep the notation simple we neglected the momentum conservation
delta functions in the higher power terms. The coefficients $\Upsilon$
and $\Xi^{k}$ in the expansion are the translation kernels originally
defined for the pure Yang-Mills theory \cite{Ettle Morris}.

\begin{eqnarray}
&&
\Upsilon_{12\cdots n}=\frac{\hat{1}\,\hat{3}\cdots\hat{n}}{\left(23\right)\left(34\right)\cdots\left(n-1,n\right)},\nonumber \\
&&
\Xi_{12\cdots n}^{k}=\frac{\hat{k}\,\hat{3}\cdots\hat{n}}{\left(23\right)\left(34\right)\cdots\left(n-1,n\right)}.\label{Upsilon and Xi}
\end{eqnarray}

The transformation expansions (\ref{SQCD 1}) to (\ref{SQCD 4}) were
verified to generate a unit Jacobian and have the effect of absorbing
the $\mathcal{L}_{A}^{++-}$, $\mathcal{L}_{\Lambda A}^{++-}$ terms
into the new lagrangian. In this paper we take a different approach
and apply the transformation on superfields directly. As noted in
section (\ref{section super Feynman rules}) the Green function can
be computed from functional integral over superfields labeled by super-space
momenta $p$ and $\eta$. Similarly a generating functional of currents
in coordinate space originally derived from integraing over physical
filed components $\mathcal{A}$, $\bar{\mathcal{A}}$, $\Lambda$,
$\bar{\Lambda}$ can be reorganised as a functional of super-currents

\begin{equation}
Z_{0}[J]=\int\mathcal{D}A\mathcal{\mathcal{D}}\bar{A}\mathcal{D}\bar{\Lambda}\mathcal{D}\Lambda \hspace{9cm} \nonumber 
\end{equation}
\begin{equation}
 exp\left\{ iS\int d^{4}x\, d\theta d\bar{\theta}\,\mathcal{L}_{free}+i\int d^{4}x\, j^{(A)}\mathcal{A}+j^{(\Lambda)}\Lambda+\bar{\mathcal{A}}j^{(\bar{A})}+\bar{\Lambda}j^{(\bar{\Lambda})}\right\} \end{equation}
\begin{equation}
=exp\left\{ \int d^{4}x\, j^{(\bar{A})}\Delta^{(A)}j^{(A)}+j^{(\bar{\Lambda})}\Delta^{(\Lambda)}j^{(\Lambda)}\right\} =exp\left\{ \int d^{4}x\, d\theta d\bar{\theta}\,\bar{J}\Delta J\right\} \end{equation}

where we defined the super-currents in coordinate space as
\begin{eqnarray}
 &&
J(x,\theta)=\theta\bar{\theta}j^{(A)}(x)-i\bar{\theta}j^{(\Lambda)},\\
&&
\bar{J}(x,\theta)=\frac{1}{i\hat{\partial}}j^{(\bar{A})}-i\theta j^{(\Lambda)},
\end{eqnarray}

and we have extracted the interaction part of the lagrangian as variation
operators with respect to the supercurrents. As in the momentum space
we introduce superfields as auxiliary field variables and restore
the full action in chiral and anti-chiral superfields by operating
the variation operators back onto the free generating functional.

\begin{equation}
Z\left[J\right]=\int\mathcal{D}\Phi(x,\theta)\mathcal{D}\bar{\Phi}(x,\theta)\, exp\left\{ iS\left[\Phi\right]+i\int J\Phi+\bar{\Phi}\bar{J}\right\} ,\end{equation}

where $S\left[\Phi\right]$ is the $\mathcal{N}=1$ SYM action in
light-cone gauge introduced in (\ref{N1 SYM lagrangian}). Inspired
by the canonical transformation originally applied on pure Yang-Mills
to derive an MHV lagrangian \cite{MHV lagrangian} we make a change
of variables. The superfield $\Phi(\tau,{\bf x},\theta)$ at light-cone
time $\tau$ is assumed to be a functional of $\chi(\tau,{\bf y},\xi)$
defined through power expansion. As in the pure Yang-Mills theory
the expansion for the anti-chiral superfield $\bar{\Phi}(\tau,{\bf x},\theta)$
is assumed to contain only one $\bar{\chi}(\tau,{\bf y},\theta)$
in each term, while the power of $\chi(\tau,{\bf y},\xi)$ increases
term by term. This arrangement ensures that the new lagrangian will
have exactly two anti-chiral superfields in every vertex as demended
by the CSW rules. We assume the transformation is given by

\begin{equation}
\bar{d}\,\bar{\Phi}^{a}(x,\theta)=\int d^{3}y\, d\xi d\bar{\xi}\frac{\delta\chi^{b}(y,\xi)}{\delta\Phi^{a}(x,\theta)}\,\bar{d}\,\bar{\chi}^{b}(y,\xi),\label{SUSY canonical transform 1}\end{equation}

and after the transformation the unwanted super-vertex is absorbed
into the new free field lagrangian $\mathcal{L}^{-+}\left[\Phi\right]+\mathcal{L}^{--+}\mbox{\ensuremath{\left[\Phi\right]}}=\mathcal{L}^{-+}\left[\chi\right]$.

We note that the nilpotency of the Grassmann variables $\theta$ $\bar{\theta}$
allows us to replace the anti-chiral superfield by $\bar{\theta}\,\bar{d}\,\bar{\Phi}$,

\begin{equation}
tr\int d^{4}x\, d\theta d\bar{\theta}\hspace{0.3cm}\bar{\Phi}\frac{\partial^{2}}{\hat{\partial}}\Phi+\bar{\Phi}[\Phi,\frac{\bar{\partial}}{\hat{\partial}}\Phi]=tr\int d^{4}x\, d\theta d\bar{\theta}\hspace{0.3cm}\bar{\theta}\,\bar{d}\bar{\Phi}\left(\frac{\partial^{2}}{\hat{\partial}}\Phi+[\Phi,\frac{\bar{\partial}}{\hat{\partial}}\Phi]\right)\label{SUSY canonical transform 2}\end{equation}

Using the condition (\ref{SUSY canonical transform 1}) and stripping
off an $\bar{\theta}\,\bar{d}\,\bar{\chi}$ from both sides of the
equation, we have

\begin{equation}
\frac{\partial^{2}}{\hat{\partial}}\Phi(x,\theta)+[\Phi,\frac{\bar{\partial}}{\hat{\partial}}\Phi](x,\theta)=\int d^{3}y\, d\xi d\bar{\xi}\,\frac{\partial^{2}}{\hat{\partial}}\chi(y,\xi)\,\frac{\delta\Phi(x,\theta)}{\delta\chi(y,\xi)}\label{SUSY canonicanl transform 3}\end{equation}

From (\ref{SUSY canonicanl transform 3}) we determine the translation
kernels in the expansions of $\Phi$ and $\bar{\Phi}$. Since the
above condition is the same as the condition used to solve for translation
kernels in the pure Yang-Mills theory \cite{MHV lagrangian}, we see
that the kernels are simply given by the same formulae as in (\ref{Upsilon and Xi}).

\begin{equation}
\Phi({\bf x}_{1},\theta)=\chi({\bf x}_{1},\theta)+\int\Upsilon_{123}\,\chi({\bf x}_{2},\theta)\,\chi({\bf x}_{3},\theta)+\cdots\label{expand phi coord space}\end{equation}
\begin{equation}
\bar{d}\,\bar{\Phi}({\bf x}_{1},\theta)=\bar{d}\,\bar{\chi}({\bf x}_{1},\theta)+\int\Xi_{123}^{2}\,\left(\bar{d}\,\bar{\chi}({\bf x}_{2},\theta)\right)\,\chi({\bf x}_{3},\theta)
\nonumber \end{equation}
\begin{equation}
\hspace{3cm} +\int\Xi_{123}^{3}\,\chi({\bf x}_{2},\theta)\,\left(\bar{d}\,\bar{\chi}({\bf x}_{3},\theta)\right)\cdots\label{expand phi bar coord space}\end{equation}

We Fourier transform the chiral superfields into momentum space and
then apply the integrals defined in equations (\ref{transform phi bar})
to (\ref{transform d phi}) to obtain the superfields in the new representation.
The expansion formulae in momentum space are
\begin{equation}
\phi(p_{1},\eta_{1})=\chi(p_{1},\eta_{1})
\hspace{9cm}
\nonumber \end{equation}
\begin{equation}
+\int\Upsilon_{123}\frac{-1}{\hat{1}^{\frac{1}{2}}}(-\eta_{1}\hat{1}^{\frac{1}{2}}+\eta_{2}\hat{2}^{\frac{1}{2}}+\eta_{3}\hat{3}^{\frac{1}{2}})\,\chi(p_{2},\eta_{2})\chi(p_{3},\eta_{3})+\cdots
\hspace{2cm}
\nonumber 
\end{equation}
\begin{equation}
+\int\Upsilon_{12\cdots n}\frac{-1}{\hat{1}^{\frac{1}{2}}}(-\eta_{1}\hat{1}^{\frac{1}{2}}+\eta_{2}\hat{2}^{\frac{1}{2}}+\cdots\eta_{n}\hat{n}^{\frac{1}{2}})\,\chi(p_{2},\eta_{2})\cdots\chi(p_{n},\eta_{n})\,+\cdots,\label{expand phi momentum space}\end{equation}

and

\begin{equation}
\bar{\phi}(p_{1},\eta_{1})=\bar{\chi}(p_{1},\eta_{1}^{'})
\hspace{9cm}
\nonumber 
\end{equation}
\begin{equation}
+\int\Xi_{123}^{2}\frac{-\hat{2}^{\frac{1}{2}}}{\hat{1}}(-\eta_{1}\hat{1}^{\frac{1}{2}}+\eta_{2}\hat{2}^{\frac{1}{2}}+\eta_{3}\hat{3}^{\frac{1}{2}})\,\bar{\chi}(p_{2},\eta_{2})\chi(p_{3},\eta_{3})+\cdots
\hspace{2cm}
\nonumber
\end{equation}
\begin{equation}
+\int\Xi_{12\cdots n}^{k}\frac{-\hat{k}^{\frac{1}{2}}}{\hat{1}}(-\eta_{1}\hat{1}^{\frac{1}{2}}+\eta_{2}\hat{2}^{\frac{1}{2}}+\cdots\eta_{n}\hat{n}^{\frac{1}{2}})\,\chi(p_{2},\eta_{2})\cdots\bar{\chi}(p_{k},\eta_{k})\cdots\chi(p_{n},\eta_{n})\,+\cdots.\label{expand phi bar momentum space}
\end{equation}

In order to avoid introducing too many symbols we slightly abuse the
notation and use $\chi$ both for superfields before and after the
integral transformations (\ref{transform phi bar}) to (\ref{transform d phi}),
in the same spirit as the same symbol is used for wave functions before
and after the Fourier transformation in the standard notation. The
distinction between these two types of fields should be clear judging
from the labels $(x,\theta)$ or $(p,\eta)$ attached to the superfields.
In equations (\ref{expand phi momentum space}) and (\ref{expand phi bar momentum space})
we neglected the overall momentum conservation deltafunction and the
integrations are understood to be performed over momenta $p_{2}$
to $p_{n}$ as well as superspace momenta $\eta_{2}$ to $\eta_{n}$.

The above expansion formulae can be conveniently summarised if we
generalise the graphical notation introduced for the pure Yang-Mills
in \cite{us 1}. When an $n$-th order term in (\ref{expand phi momentum space})
contribute to the calculation we use a blank circle follow by $(n+1)$
lines to represent the translation kernel, where one of the lines
comes from the superfield $\phi$ being translated. For the $\bar{\phi}$
translation, we use a similar graph with the blank circle replaced
by a gray circle.


\begin{figure}[!h]
  \centering 
  \subfigure{
  \begin{picture}(108,64) (9,7)
    \SetWidth{0.375}

    \Line(15,34)(36,34)

    \Line(45,34)(72,13)
    \Line(45,34)(75,52)
    \Text(12,31)[br]{$\phi$}
    
    \Text(85,54)[br]{$ \chi $}
    \Text(85,37)[br]{.}
    \Text(85,34)[br]{.}
    \Text(85,31)[br]{.}
    \Text(85,28)[br]{.}
    \Text(85,3)[br]{$ \chi$}
       \BCirc(45,34){8}
  \end{picture}}
  \subfigure{
  \begin{picture}(108,64) (9,7)
    \SetWidth{0.375}

    \Line(15,34)(36,34)

    \Line(45,34)(72,13)
    \Line(45,34)(75,52)
    \Line(48,34)(78,34)
    \Text(12,31)[br]{$\bar{\phi} $}
    \Text(85,44)[br]{.}
    \Text(85,47)[br]{.}
    \Text(85,41)[br]{.}
    
    \Text(85,54)[br]{$ \chi $}
    
    \Text(85,21)[br]{.}
    \Text(85,18)[br]{.}
    \Text(85,15)[br]{.}
    \Text(85,27)[br]{$ \bar{\chi}$}
    \Text(85,3)[br]{$ \chi $}
       \GCirc(45,34){8}{0.75}
           \Vertex(58,34){2}
  \end{picture}}
  \caption{Graphical representations of superfield expansions}
  \label{N1 kernel}
  \end{figure}
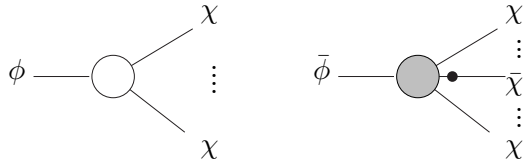

\subsection{Generating MHV super-vertices}

\label{section N1 MHV super vertices}

Following the same steps as for the pure Yang-Mills theory the super-amplitude
given by applying LSZ reduction to the Green function is generically
transformed into a series, each of the term contains a number of translation
kernels

\begin{eqnarray}
&&
\prod_{l}p_{l}^{2}\left\langle \cdots\phi_{i}\cdots\bar{\phi}_{j}\cdots\right\rangle  \nonumber \\
&&
=\sum_{m,n}\prod_{l}p_{l}^{2} 
\left\langle \cdots\left(\Upsilon_{i\, i_{2}\cdots i_{m}}\chi_{i_{2}}\chi_{i_{3}}\cdots\chi_{i_{m}}\right)\cdots\left(\Xi_{j\, j_{2}\cdots j_{n}}^{k}\chi_{j_{2}}\cdots\bar{\chi}_{j_{k}}\cdots\chi_{j_{n}}\right)\cdots\right\rangle .
\end{eqnarray}

At tree-level these kernels are supressed by LSZ facotrs $p_{l}^{2}$,
so an MHV super-amplitude simply equals the on-shell limit of the
corresponding MHV super-vertex. In section (\ref{N1 BCFW}) we derived
the formula for a generic n-point $\mathcal{N}=1$ MHV super-amplitude
from supersymmetry BCFW recursion, which can only differ from the
MHV super-vertices by squares of leg momenta. Since the expansion
coefficients in (\ref{expand phi momentum space}) and (\ref{expand phi bar momentum space})
are holomorphic, such difference is absent, so an n-point MHV super-vertex
is provided by the same formula as for the super-amplitude.

Alternatively, we can directly compute the n-point MHV super-vertex.
An MHV super-vertex in the new lagrangian receives contributions from
the original 3-point (\ref{3pt N1 MHV}) and the 4-point vertices
(\ref{4pt N1 1}), (\ref{4pt N1 2}), and the superfields attached
to the legs of the vertices branch into trees of new field variables
according to the expansion formulae (\ref{expand phi momentum space})
and (\ref{expand phi bar momentum space}). In the appendix \ref{appendix proof}
we prove that the formula for an n-point MHV super-vertex is the same
as the super-amplitude (\ref{N1 MHV supervertex}) by matching their
coefficients under partial fraction expansion. 

\begin{equation}
V(1^{+},2^{+}\dots i^{-},j^{-},\dots n^{+})=\frac{\left\langle i\, j\right\rangle ^{3}}{\left\langle 12\right\rangle \left\langle 23\right\rangle \cdots\left\langle n1\right\rangle }\left(\sum_{a,b=1}^{n}\left\langle a\, b\right\rangle \eta_{a}\eta_{b}\right).\label{N1 MHV supervertex}\end{equation}

As noted in section (\ref{section super Feynman rules}), although
$\phi$ and $\bar{\phi}$ are introduced as integration variables,
from current algebra the superfields appear in the Green function
can be interchanged by the corresponding combinations of gluon and
gluino fields. Substituting (\ref{phi}) and (\ref{phi bar}) into
the canonical transformation formulae of superfields (\ref{expand phi coord space})
and (\ref{expand phi bar coord space}) suggests that the MHV lagrangian
can be as well derived by transforming physical field components seperately.
Writing the new field variables as
\begin{eqnarray}
&&
\chi(x,\theta)=\mathcal{B}(y)+i\theta\,\Pi(y), \\
&&
\bar{\chi}(x,\theta)=\overline{\mathcal{B}(y)}+i\bar{\theta}\,\overline{\Pi(y)},
\end{eqnarray}

and integrating over Grassmann variables $\theta$ and $\bar{\theta}$,
we find the same transformation relations as equations (\ref{SQCD 1})
to (\ref{SQCD 4}) originally given by Morris and Xiao in \cite{Xiao SQCD}.

\subsection{SUSY Ward identity}

\label{section SUSY Ward}

In \cite{twistor string} Witten introduced an on-shell representation
of the SUSY generators for $\mathcal{N}=4$ SYM theory and verified
that the MHV super-amplitude given by Nair \cite{Nair} is superconformal
invariant. We find both the on-shell SUSY generators and the super-amplitude
for $\mathcal{N}=4$ theory resemble the formulae we derived in (\ref{onshell SUSY gen})
and (\ref{N1 MHV supervertex}). It is then straightforward to verify
that the n-point $\mathcal{N}=1$ MHV super-amplitude is SUSY invariant.
The on-shell SUSY transformation operator $Q(\xi)$ is given by

\begin{equation}
Q(\xi)=\sum_{i}^{n}\left\langle \xi i\right\rangle \eta_{i}+\left[i\xi\right]\frac{\partial}{\partial\eta_{i}}\end{equation}

The transformation operator consists of a multiplication part and
a differentiation part. When operating on formula (\ref{N1 MHV supervertex})
we find the two parts are separately zero. Collecting terms having
the same Grassmann numbers, the multiplication part vanishes because
from Jacobi identity

\begin{equation}
\left\langle \xi1\right\rangle \left\langle 23\right\rangle \eta_{1}\eta_{2}\eta_{3}+\left\langle \xi3\right\rangle \left\langle 12\right\rangle \eta_{3}\eta_{1}\eta_{2}+\left\langle \xi2\right\rangle \left\langle 31\right\rangle \eta_{2}\eta_{3}\eta_{1}=0\end{equation}

for any three of the momenta carried by external lines. The differentiation
part is proportional to

\begin{equation}
\sum_{i}\left[\xi i\right]\left\langle ij\right\rangle \eta_{j}=0\end{equation}

which vanishes from conservation of momentum. We note that supersymmetry
is taken as a build-in property of the super-amplitude. Since the
functional integral is invariant under SUSY transformation 

\begin{equation}
\prod_{i}p_{i}^{2}\left\langle Q(\xi)\phi_{1}\phi_{2}\bar{\phi}_{3}\cdots\bar{\phi}_{n}\right\rangle =0.\label{SUSY Ward id}\end{equation}

Differentiating both sides of the identity (\ref{SUSY Ward id}) with
respect to $\eta_{1}$ $\eta_{2}$ $\eta_{j}$ for example, gives
the familiar SUSY Ward identity relating the amplitude that consists
of a pair of gluino and the amplitude of all gluons.

\begin{equation}
\left\langle 21\right\rangle \left\langle 1_{\Lambda}^{+},2^{+},3^{-},\cdots j_{\Lambda}^{-}\cdots n^{-}\right\rangle +\left\langle 2j\right\rangle \left\langle 1^{+},2^{+},3^{-},\cdots n^{-}\right\rangle =0\end{equation}

\section{Conclusion and discussions}

In this paper we performed a fermionic integral transformation followed
by a supersymmetric canonical transformation on the superfield variables
to directly generate Nair's like $\mathcal{N}=1$ MHV super-vertices 
(\ref{N1 MHV supervertex}). We showed that in such a single Grassmann variable
representation the superfields are unconstrained and act as functional integration
variables. The supersymmetric Feynman rules derived from the functional integral 
provide the $\mathcal{N}=1$ version CSW rules, in which the reference spinors
used to extract spinors of internal lines  
are all chosen as $(0,1)^{T}$, as in the case of pure Yang-Mills theory \cite{MHV lagrangian}. 
(Other options correspond to
using different null-vectors to define light-cone coordinates.) We applied BCFW recursion
to compute the generic n-point helicity-ordered MHV super-amplitude defined through 
superfield Green functions. Symmetrising with respect to all helicity configurations
this reproduces the super-amplitude formula derived in \cite{Bern:2009xq}.

As an alternative approach we note that the MHV lagrangian can be derived entirely from
 the spinor language. In this approach the gauge field and gaugino field components are
 spanned by the bispinors and spinors consisting of various combinations of $\mu_{\alpha}$ and $\lambda_{\alpha}$, 
where 

\begin{eqnarray}
&&
\mu_{\alpha}=\left(\begin{array}{c}
\sqrt{\check{p}-p\bar{p}/\hat{p}}\\
0\end{array}\right),\,\bar{\mu}_{\dot{\alpha}}=\left(\begin{array}{c}
\sqrt{\check{p}-p\bar{p}/\hat{p}}\\
0\end{array}\right),\label{spinors eta}
\\
 &&
\lambda_{\alpha}=\left(\begin{array}{c}
-p/\sqrt{\hat{p}}\\
\sqrt{\hat{p}}\end{array}\right),\hspace{1cm} \bar{\lambda}_{\dot{\alpha}}=\left(\begin{array}{c}
-\bar{p}/\sqrt{\hat{p}}\\
\sqrt{\hat{p}}\end{array}\right),\label{spinor lambda} 
\end{eqnarray}

and a generic off-shell momentum is written as  $P_{\alpha\dot{\alpha}}=\mu_{\alpha}\bar{\mu}_{\dot{\alpha}}+\lambda_{\alpha}\bar{\lambda}_{\dot{\alpha}}$. We identify the
 positive, negative helicity physical field components as the coefficients of the terms $\eta_{\alpha}\bar{\lambda}_{\dot{\alpha}}$, $\lambda_{\alpha}\bar{\eta}_{\dot{\alpha}}$ ,$\lambda_{\alpha}$, $\bar{\lambda}_{\dot{\alpha}}$ respectively. Analogously 
to deriving the light-cone gauge Yang-Mills lagrangian we eliminate the $\lambda_{\alpha} \bar{\lambda}_{\dot{\alpha}}$ component of the gluon field 
by fixing gauge condition and integrate over all unphysical components and Grassmann variables $\langle\theta \mu\rangle$, $[\theta \mu]$. The 
resulting lagrangian is the same as the one built from chiral superfields which are composed of helicity field components and Grassmann variables $\langle\theta\lambda\rangle$, $[\theta\lambda]$.
From this point 
of veiw the fermionic integral transformation defined through equations (\ref{transform phi bar}) to (\ref{transform d phi})
 can be regarded as the off-shell version of identifying $\eta_{A}= [\lambda \theta_{A} ]$ in the momentum twistor space \cite{Mason:2009qx, Ferber:1977qx} and the chiral prescription described above is the momentum space analogue to the fields developed in \cite{Boels:2006ir} 
by Boels, Mason and Skinner in the ordinary twistor space.

With a few modifications the method can be extended to $\mathcal{N}=4$
SYM theory. It has long been known that the $\mathcal{N}=4$ SYM lagrangian in
 light-cone gauge can be constructed by chiral superfields \cite{Brink Ramond}. 
The corresponding fermionic integral transformation which gives rise to
 the Grassmann variables $\eta_{A}$ in $\mathcal{N}=4$ SYM theory 
(and notably in $\mathcal{N}=8$ supergravity) was found by Kallosh in \cite{Kallosh:2009db}. 
The generic n-point vertices in the $\mathcal{N}=4$ MHV lagrangian can be shown to assume the form
described by Nair's formula \cite{Nair} either from holomrphy or from explicit computation.

\section*{Acknowledgements}
We would like to thank Paul Mansfield for comments and for bringing our attention to recent developments in twistor field theory. We are also grateful for the discussions with Paolo Benincasa, James Ettle and Zhiguang Xiao.

\appendix
\section{Notation}

\label{appendix notation}

Here we summarise the notation used throughout this paper.

As a shorthand notation we use $\left(\check{p},\,\hat{p},\, p,\,\bar{p}\right)$
to describe covariant vectors in the light-cone coordinates, which
is related to the Minkowski coordinates by

\begin{equation}
\check{p}=\left(p_{0}-p_{3}\right),\,\hat{p}=\left(p_{0}+p_{3}\right),\, p=\left(p_{1}-ip_{2}\right),\,\bar{p}=\left(p_{1}+ip_{2}\right).\end{equation}

In the light-cone coordinates the metric becomes off-diagonal. The
Lorentz invariant product of two vectors is given by

\begin{equation}
p\cdot q=\left(\check{p}\hat{q}+\hat{p}\check{q}-p\bar{q}-\bar{p}q\right)/2.\end{equation}

To keep the notation as simple as possible, the momentum components
$p_{n\,\mu}$ of the $n^{th}$ external leg are denoted by number
$n$ with the appropriate decoration $\left(\check{n},\,\hat{n},\,\tilde{n},\,\bar{n}\right)$.
Note that a tilde is used for the $p=p_{1}-ip_{2}$ component to avoid
possible confusions with numerical factors.

A 4-vector can be written in the form of a bispinor 

\begin{equation}
P_{\alpha\dot{\alpha}}=p^{\mu}\sigma_{\mu\alpha\dot{\alpha}}=\left(\begin{array}{cc}
\check{p} & -p\\
-\bar{p} & \hat{p}\end{array}\right)\end{equation}

by contracting with $\sigma_{\mu}=\left(I_{2},\,\vec{\sigma}\right)$,
where $I_{2}$ is the $2\times2$ identity matrix and $\vec{\sigma}$
stands for Pauli matrices.

If $p^{\mu}$ is lightlike, $\check{p}=p\bar{p}/\hat{p}$ and the
bispinor factorises $p_{\alpha\dot{\alpha}}=\lambda_{\alpha}\bar{\lambda}_{\dot{\alpha}}$,
where

\begin{equation}
\lambda_{\alpha}=\left(\begin{array}{c}
-p/\sqrt{\hat{p}}\\
\sqrt{\hat{p}}\end{array}\right),\,\bar{\lambda}_{\dot{\alpha}}=\left(\begin{array}{c}
-\bar{p}/\sqrt{\hat{p}}\\
\sqrt{\hat{p}}\end{array}\right).\end{equation}

Spinors $\lambda_{i\alpha}$ associated with different massless particles
can be contracted to given a Lorentz invariant angle bracket

\begin{equation}
\left\langle 12\right\rangle =\epsilon^{\alpha\beta}\lambda_{1\alpha}\lambda_{2\beta}=\frac{\left(12\right)}{\sqrt{\hat{1}\hat{2}}},\end{equation}

and we define a round bracket as 

\begin{equation}
\left(12\right)=\hat{1}\tilde{2}-\hat{2}\tilde{1}.\end{equation}

\section{$\mathcal{N} = 1$ MHV super-vertices}
\label{appendix proof}

In this appendix we prove that when the superfields $\phi$ and $\bar{\phi}$
are canonically transformed into new fields $\chi$ and $\bar{\chi}$
the original 3-point and 4-point vertices in the light-cone gauge
SYM generate MHV super-vertices of the form (\ref{N1 MHV supervertex}).
For simplicity we show this is true when the two negative helicity
particles are adjacent. The method outlined here generalise to arbitrary
configurations.

Labeling the negative helicity leg momenta as leg $1$ and $2$ the
formula (\ref{N1 MHV supervertex}) reads 

\begin{equation}
V(1^{-},2^{-},3^{+}\dots n^{+})=\frac{\left\langle 1\,2\right\rangle ^{3}}{\left\langle 12\right\rangle \left\langle 23\right\rangle \cdots\left\langle n1\right\rangle }\left(\sum_{a,b=1}^{n}\left\langle a\, b\right\rangle \eta_{a}\eta_{b}\right)
\hspace{2cm}
\end{equation}
\begin{equation}
\hspace{1.4cm}
=\frac{\left(12\right)^{2}}{\left(23\right)\cdots\left(n1\right)}\,\frac{\hat{3}\cdots\hat{n}}{\hat{1}^{\frac{1}{2}}\hat{2}^{\frac{1}{2}}}\left(\sum_{a,b=1}^{n}\frac{\left(a\, b\right)}{\hat{a}^{\frac{1}{2}}\hat{b}^{\frac{1}{2}}}\eta_{a}\eta_{b}\right)\label{N1 Nair formula 2}
\end{equation}

\begin{figure}[!h]
  \centering 
  \subfigure{
    \begin{picture}(0,5) 
    \end{picture}
  }
  \\
    \subfigure{

  \begin{picture}(81,78) (24,-9)
    \SetWidth{0.375}

    \Line(45,48)(84,12)
    \Line(45,12)(84,48)

    \Line(87,57)(90,69)
    \Line(93,51)(105,54)
    \Line(24,6)(36,9)
    \Line(39,-9)(42,3)
    \Line(42,57)(39,69)
    \Line(36,48)(24,51)
    \Line(90,-9)(87,3)
    \Line(105,6)(93,9)
    \SetWidth{0.5}
    \Vertex(59,25){2}
    \Vertex(59,35){2}
    \Vertex(64,30){4}
    \SetWidth{0.375}
    \Line(28,1)(37,7)

    \GCirc(44,49){8}{0.75}
    \GCirc(44,11){8}{0.75}    
    \BCirc(85,11){8}
    \BCirc(85,49){8}
    
        \Text(22,3)[br]{$1$}
        \Text(27,-6)[br]{$n$}
        \Text(50,-20)[br]{$l+1$}
        \Text(30,-6)[br]{.}
        \Text(33,-9)[br]{.}
                \Text(95,-20)[br]{$l$}
                \Text(135,3)[br]{$m+1$}
                \Text(110,-3)[br]{.}
                \Text(107,-6)[br]{.}
                \Text(104,-9)[br]{.}
                          \Text(112,73)[br]{$k+1$}
                          \Text(117,55)[br]{$m$}
                          \Text(110,61)[br]{.}
                          \Text(107,64)[br]{.}
                          \Text(104,67)[br]{.}
                                \Text(40,73)[br]{$k$}
                                \Text(21,52)[br]{$2$}
                                \Text(27,61)[br]{.}
                                \Text(30,64)[br]{.}
                                \Text(33,67)[br]{.}
        \Vertex(30,49){2}
         \Vertex(30,8){2}
  \end{picture} 
    
  }   \caption{Translated 4-point vertex}
\label{spider graph}
  \end{figure}
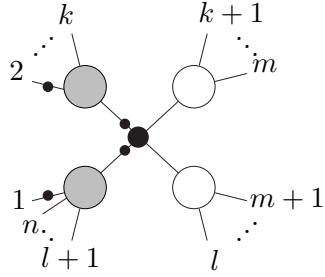

\begin{figure}[!h]
\centering
  \subfigure{
    \begin{picture}(0,0) (10,10)
    \end{picture}
  }
  \\
  \subfigure{
    \begin{picture}(81,78) (24,-9)
    \SetWidth{0.375}

    \Line(45,48)(64,30)
        \Line(64,30)(85,30)
    \Line(45,12)(64,30)

    \Line(90,34)(102,42)
    \Line(24,6)(36,9)
    \Line(39,-9)(42,3)
    \Line(42,57)(39,69)
    \Line(36,48)(24,51)

    \Line(102,18)(90,26)
    \SetWidth{0.5}
    \Vertex(59,25){2}
    \Vertex(59,35){2}
    \Vertex(64,30){4}
    \SetWidth{0.375}
    \Line(28,1)(37,7)

    \GCirc(44,49){8}{0.75}
    \GCirc(44,11){8}{0.75}    
    \BCirc(85,30){8}
    
        \Text(22,3)[br]{$1$}
        \Text(27,-6)[br]{$n$}
        \Text(50,-20)[br]{$l+1$}
        \Text(30,-6)[br]{.}
        \Text(33,-9)[br]{.}
                \Text(125,47)[br]{$k+1$}
                \Text(110,13)[br]{$l$}
                \Text(110,33)[br]{.}
                \Text(110,30)[br]{.}
                \Text(110,27)[br]{.}

                                \Text(40,73)[br]{$k$}
                                \Text(21,52)[br]{$2$}
                                \Text(27,61)[br]{.}
                                \Text(30,64)[br]{.}
                                \Text(33,67)[br]{.}
        \Vertex(30,49){2}
        \Vertex(30,8){2}
  \end{picture}
  
  }  \caption{Translated 3-point vertex} \label{lobster graph}
\end{figure}
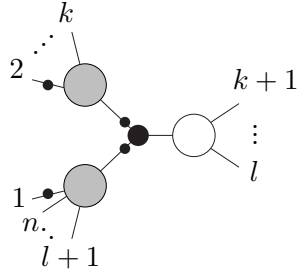

From (\ref{Upsilon and Xi}) we saw the factors $\Upsilon$ and $\Xi$
appearing in the translation formula (\ref{expand phi momentum space})
and (\ref{expand phi bar momentum space}) contain in the denominators
a sequential product of round brackets. After the canonical transformation
the 3-point and 4-point vertices constitute three and four groups
of products of round brackets (Fig.\ref{spider graph} and \ref{lobster graph}).
Therefore when regarded as functions of tilde component variables
$\tilde{p}$ which are contained in the round brackets, both formula
(\ref{N1 Nair formula 2}) and the translated 3-point and 4-point
vertices can be spanned by terms of the form
\begin{eqnarray}
&&\frac{1}{\left(23\right)\left(34\right)\cdots\left(k-1,k\right)}\times\frac{1}{\left(k+1,k+2\right)\cdots\left(m-1,m\right)}
\nonumber
\\
&&
\times\frac{1}{\left(m+1,m+2\right)\cdots\left(l-1,l\right)}\times\frac{1}{\left(l+1,l+2\right)\cdots\left(n,1\right)}\label{partial fraction4}\end{eqnarray}

together with terms of the form
\begin{equation}
\frac{\left(12\right)}{\left(23\right)\left(34\right)\cdots\left(k-1,k\right)}\times\frac{1}{\left(k+1,k+2\right)\cdots\left(l-1,l\right)}\times\frac{1}{\left(l+1,l+2\right)\cdots\left(n,1\right)}\label{partial fraction3}\end{equation}

where the expansion coefficients depend on pairs of Grassmann numbers
$\eta_{i}\eta_{j}$ with $i$ $j$ running through all possible combinations
of external legs and on hat component momenta $\hat{p}$. 

To obtain the coefficients we use the method of partial fractions.
For terms of the form (\ref{partial fraction4}) we adjust tilde components
to set 
\begin{eqnarray}
&&
\left(12\right)=0 \\
&&
\left(23\right)=\cdots=\left(k-1,k\right)=0 \\
&&
\left(k+1,k+2\right)=\cdots=\left(m-1,m\right)=0 \\
&&
\left(m+1,m+2\right)=\cdots=\left(l-1,l\right)=0 \\
&&
\left(l+1,l+2\right)=\cdots=\left(n,1\right)=0\label{condition 4}
\end{eqnarray}

Similarly, the coefficient of term (\ref{partial fraction3}) can
be obtained by applying the conditions
\begin{eqnarray}
 &&
\left(23\right)=\cdots=\left(k-1,k\right)=0 \\
&&
\left(k+1,k+2\right)=\cdots=\left(l-1,l\right)=0 \\
&&
\left(l+1,l+2\right)=\cdots=\left(n,1\right)=0\label{condition 3}
\end{eqnarray}

We shall prove that the translated 3-point and 4-point vertices contribute
to give the formula (\ref{N1 Nair formula 2}) by showing their expansion
coefficients agree with each other.

First we check the coefficients for four sequential products of brackets
(\ref{partial fraction4}). Applying conditions (\ref{condition 4})
on formula (\ref{N1 Nair formula 2}) gives zero because of the $\left(12\right)$
dependence in the numerator. The 3-point vertex contribution to the
coefficient of (\ref{partial fraction4}) can be from the following
three cases

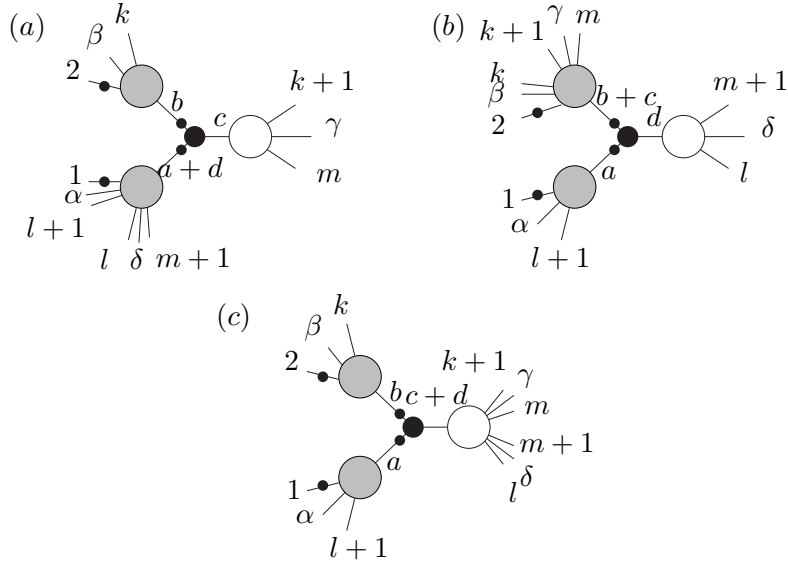
\begin{figure}[!h]
  \centering 
  \subfigure{
    \begin{picture}(5,5) 
    \end{picture}
  }
  \\
  \subfigure{
    \begin{picture}(5,78) 
                \Text(0,75)[br]{$(a)$}
    \end{picture}
  }
    \subfigure{
      \begin{picture}(81,78) (24,-9)
    \SetWidth{0.375}

    \Line(45,48)(64,30)
        \Line(64,30)(85,30)
    \Line(45,12)(64,30)

    \Line(90,34)(102,42)
    \Line(24,13)(36,13)
    \Line(39,-9)(42,3)
    \Line(42,57)(39,69)
        \Line(40,50)(32,60) 
    \Line(36,48)(24,51)


    \Line(102,18)(90,26)
          \Line(108,30)(90,30) 
    \SetWidth{0.5}
    \Vertex(59,25){2}
    \Vertex(59,35){2}
    \Vertex(64,30){4}
         \Text(75,14)[br]{$a+d$}
         \Text(60,40)[br]{$b$}
         \Text(76,34)[br]{$c$}
    \SetWidth{0.375}
    \Line(25,4)(37,8)
    \Line(47,-8)(46,5)    
    \Line(43,-10)(44,5)  
    \Line(23,8)(37,10)  

    \GCirc(44,49){8}{0.75}
    \GCirc(44,11){8}{0.75}    
    \BCirc(85,30){8}
    
        \Text(22,12)[br]{$1$}
        \Text(22,5)[br]{$\alpha$}        
        \Text(23,-9)[br]{$l+1$}
        \Text(32,-20)[br]{$l$}
        \Text(45,-20)[br]{$\delta$}
        \Text(78,-20)[br]{$m+1$}
                \Text(125,47)[br]{$k+1$}
                 \Text(120,30)[br]{$\gamma$}
                \Text(120,13)[br]{$m$}

                                \Text(40,73)[br]{$k$}
                                 \Text(30,63)[br]{$\beta$}
                                \Text(21,52)[br]{$2$}
           \Vertex(30,49){2}
            \Vertex(30,13){2}
  \end{picture}    
}
  \subfigure{
    \begin{picture}(40,0) 
    \end{picture}
  }
  \subfigure{
    \begin{picture}(10,78) 
                \Text(0,75)[br]{$(b)$}
    \end{picture}
  }   
 \subfigure{
      \begin{picture}(81,78) (24,-9)
    \SetWidth{0.375}

    \Line(45,48)(64,30)
        \Line(64,30)(85,30)
    \Line(45,12)(64,30)

    \Line(90,34)(102,42)
    \Line(24,6)(36,9)
    \Line(39,-9)(42,3)
      \Line(30,-3)(44,11)  
    \Line(45,57)(46,69)  
    \Line(39,42)(24,37)  
              \Line(108,30)(90,30) 
                \Line(40,68)(44,49)  
        \Line(24,50)(44,48)  
          \Line(24,46)(44,46)  
        \Line(34,63)(40,54)

    \Line(102,18)(90,26)
    \SetWidth{0.5}
    \Vertex(59,25){2}
    \Vertex(59,35){2}
    \Vertex(64,30){4}
         \Text(75,41)[br]{$b+c$}
         \Text(60,14)[br]{$a$}
         \Text(77,34)[br]{$d$}
    \SetWidth{0.375}

    \GCirc(44,49){8}{0.75}
    \GCirc(44,11){8}{0.75}    
    \BCirc(85,30){8}
    
        \Text(22,3)[br]{$1$}
               \Text(27,-6)[br]{$\alpha$}
        \Text(50,-20)[br]{$l+1$}
                \Text(125,47)[br]{$m+1$}
                            \Text(120,30)[br]{$\delta$}
                \Text(110,13)[br]{$l$}

                                \Text(54,73)[br]{$m$}
                                 \Text(40,73)[br]{$\gamma$}
                                \Text(18,31)[br]{$2$}
                                 \Text(18,41)[br]{$\beta$}
                                    \Text(18,51)[br]{$k$}
                                    \Text(33,66)[br]{$k+1$}
          \Vertex(30,39){2}
           \Vertex(30,8){2}
  \end{picture}    
}
\\
  \subfigure{
    \begin{picture}(5,10) 
    \end{picture}
  }
  \\
  \subfigure{
    \begin{picture}(10,78) 
        \Text(0,75)[br]{$(c)$}
    \end{picture}
  } 
   \subfigure{
      \begin{picture}(81,78) (24,-9)
    \SetWidth{0.375}

    \Line(45,48)(64,30)
        \Line(64,30)(85,30)
    \Line(45,12)(64,30)

          \Line(30,-3)(44,11)  
          \Line(40,50)(32,60) 
    \Line(90,33)(102,42)  
     \Line(90,34)(98,44)  
    \Line(24,6)(36,9)
    \Line(39,-9)(42,3)
    \Line(42,57)(39,69)
    \Line(36,48)(24,51)

        \Line(90,28)(102,23)  
        \Line(90,32)(102,36)  

    \Line(102,18)(90,27)  
     \Line(98,16)(90,26)  
    \SetWidth{0.5}
    \Vertex(59,25){2}
    \Vertex(59,35){2}
    \Vertex(64,30){4}

    \SetWidth{0.375}
    
    \GCirc(44,49){8}{0.75}
    \GCirc(44,11){8}{0.75}    
    \BCirc(85,30){8}
    
        \Text(22,3)[br]{$1$}
                   
        \Text(55,-20)[br]{$l+1$}
                \Text(100,51)[br]{$k+1$}
                  \Text(110,45)[br]{$\gamma$}
                    \Text(133,20)[br]{$m+1$}
                    \Text(116,35)[br]{$m$}
                \Text(110,8)[br]{$\delta$}
                 \Text(104,2)[br]{$l$}

                                \Text(40,73)[br]{$k$}
                                \Text(21,52)[br]{$2$}
                                                \Text(30,63)[br]{$\beta$}
                                              \Text(27,-6)[br]{$\alpha$}
         \Vertex(30,49){2}
          \Vertex(30,8){2}
          \Text(60,14)[br]{$a$}
         \Text(60,40)[br]{$b$}
         \Text(85,38)[br]{$c+d$}
  \end{picture}    
}
  \label{lobsters abc}
\end{figure}

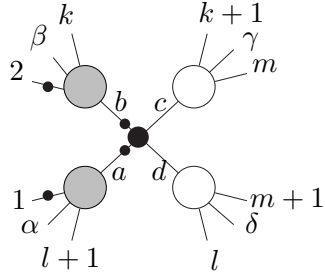
\begin{figure}[!h]
  \centering 
  \subfigure{
    \begin{picture}(0,5) 
    \end{picture}
  }
  \\
    \subfigure{

  \begin{picture}(81,78) (24,-9)
    \SetWidth{0.375}

    \Line(45,48)(84,12)
    \Line(45,12)(84,48)

          \Line(30,-3)(44,11)  
          \Line(40,50)(32,60) 
          \Line(100,63)(85,49)  
          \Line(85,10)(100,-3) 

    \Line(87,57)(90,69)
    \Line(93,51)(105,54)
    \Line(24,6)(36,9)
    \Line(39,-9)(42,3)
    \Line(42,57)(39,69)
    \Line(36,48)(24,51)
    \Line(90,-9)(87,3)
    \Line(105,6)(93,9)
    \SetWidth{0.5}
    \Vertex(59,25){2}
    \Vertex(59,35){2}
    \Vertex(64,30){4}
    \SetWidth{0.375}

    \GCirc(44,49){8}{0.75}
    \GCirc(44,11){8}{0.75}    
    \BCirc(85,11){8}
    \BCirc(85,49){8}
    
        \Text(22,3)[br]{$1$}
        \Text(50,-20)[br]{$l+1$}
                \Text(95,-20)[br]{$l$}
                \Text(135,3)[br]{$m+1$}
                          \Text(112,73)[br]{$k+1$}
                          \Text(117,55)[br]{$m$}
                                \Text(40,73)[br]{$k$}
                                \Text(21,52)[br]{$2$}
        \Vertex(30,49){2}
         \Vertex(30,8){2}
            \Text(60,14)[br]{$a$}
            \Text(60,40)[br]{$b$}
                  \Text(75,40)[br]{$c$}
                  \Text(75,14)[br]{$d$}
                 \Text(30,63)[br]{$\beta$}
                 \Text(27,-6)[br]{$\alpha$}
                  \Text(110,63)[br]{$\gamma$}
                 \Text(110,-6)[br]{$\delta$}
  \end{picture} 
    
  }   \caption{Notation for the 4-point vertex expansion}
\label{4 point notation}
  \end{figure}

In each graph one of the sequential products of brackets splits into
two. Summing over contributions from these three graphs gives

\begin{equation}
\sum_{\alpha,\beta,\gamma,\delta}\frac{1}{c^{\frac{1}{2}}d^{\frac{1}{2}}\left(b+c\right)^{2}}\left(\left(ac+bd\right)\hat{\alpha}^{\frac{1}{2}}\hat{\beta}^{\frac{1}{2}}\eta_{\alpha}\eta_{\beta}-a\left(b+c\right)\hat{\beta}^{\frac{1}{2}}\hat{\gamma}^{\frac{1}{2}}\eta_{\beta}\eta_{\gamma}-b\left(a+d\right)\hat{\delta}^{\frac{1}{2}}\hat{\alpha}^{\frac{1}{2}}\eta_{\delta}\eta_{\alpha}\right.
\nonumber
\end{equation}
\begin{equation}
\hspace{3cm} \left. +2\, ab\hat{\gamma}^{\frac{1}{2}}\hat{\delta}^{\frac{1}{2}}\eta_{\gamma}\eta_{\delta}+b\left(a+d\right)\hat{\alpha}^{\frac{1}{2}}\hat{\gamma}^{\frac{1}{2}}\eta_{\alpha}\eta_{\gamma}+a\left(b+c\right)\hat{\beta}^{\frac{1}{2}}\hat{\delta}^{\frac{1}{2}}\eta_{\beta}\eta_{\delta}\right)\label{3 point coefficient}\end{equation}

where we used $\eta_{\alpha}$ $\eta_{\beta}$ $\eta_{\gamma}$ $\eta_{\delta}$
to denote Grassmann variables associated with legs from each of the
four branches emerging from the original 4-point vertex. The index
$\alpha$ is to be summed over from $(l+1)$ to $1$, $\beta$ from
$2$ to $k$, $\gamma$ from $(k+1)$ to $m$, and finally $\delta$
from $(m+1)$ to $l$. For simplicity we denote the four internal
lines of (Fig.\ref{4 point notation}) by $a$, $b$, $c$ and $d$.
\begin{eqnarray}
&&
a=\widehat{l+1}+\cdots+\hat{n}+\hat{1} \\
&&
b=\hat{2}+\hat{3}+\cdots+\hat{k} \\
&&
c=\widehat{k+1}+\cdots+\hat{m} \\
&&
d=\widehat{m+1}+\cdots+\hat{l}
\end{eqnarray}

The contribution from the original 4-point vertex can be readily derived
by translating the superfields $\phi$ and $\bar{\phi}$ attached
to (\ref{4pt N1 1}), which cancels (\ref{3 point coefficient}).
The expansion coefficient of terms (\ref{partial fraction4}) vanishes,
therefore agrees with the coefficients obtained by expanding the formula
(\ref{N1 Nair formula 2})

\begin{figure}[!h]
\centering
  \subfigure{
    \begin{picture}(0,0) (10,10)
    \end{picture}
  }
  \\
  \subfigure{
    \begin{picture}(81,78) (24,-9)
    \SetWidth{0.375}

    \Line(45,48)(64,30)
        \Line(64,30)(85,30)
    \Line(45,12)(64,30)

    \Line(90,34)(102,42)
    \Line(24,6)(36,9)
    \Line(39,-9)(42,3)
    \Line(42,57)(39,69)
    \Line(36,48)(24,51)
          \Line(30,-3)(44,11)  
          \Line(40,50)(32,60) 
          \Line(108,30)(90,30) 
    \Line(102,18)(90,26)
    \SetWidth{0.5}
    \Vertex(59,25){2}
    \Vertex(59,35){2}
    \Vertex(64,30){4}
    \SetWidth{0.375}

    \GCirc(44,49){8}{0.75}
    \GCirc(44,11){8}{0.75}    
    \BCirc(85,30){8}
    
        \Text(22,3)[br]{$1$}
        \Text(50,-20)[br]{$l+1$}
                \Text(125,47)[br]{$k+1$}
                \Text(110,13)[br]{$l$}

                                \Text(40,73)[br]{$k$}
                                \Text(21,52)[br]{$2$}
        \Vertex(30,49){2}
        \Vertex(30,8){2}
          \Text(60,14)[br]{$a$}
         \Text(60,40)[br]{$b$}
         \Text(75,38)[br]{$c$}
                 \Text(30,63)[br]{$\beta$}
                 \Text(27,-6)[br]{$\alpha$}
                 \Text(120,30)[br]{$\gamma$}
  \end{picture}
  
  }  \caption{Notation for the 3-point vertex expansion} 
  \label{3 point notation}
\end{figure}
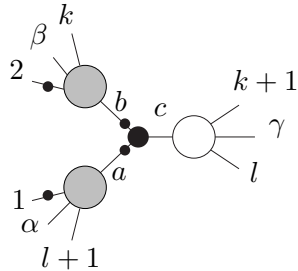

The coefficients of term (\ref{partial fraction3}) can be calculated
using the same method. However we note that since in the pure YM case
the LCYM 3-point vertex was verified to give the same expansion coefficients
as the Parke-Taylor formula, which corresponds to the $\eta_{1}\eta_{2}$
term of the formula (\ref{N1 Nair formula 2}), the proof is complete
as long as the ratio between the expansion coefficients of each $\eta_{i}\eta_{j}$
term from the original 3-point vertex is the same as ratio of coefficients
of $\eta_{i}\eta_{j}$ from the formula (\ref{N1 Nair formula 2}).
Using $\alpha$, $\beta$ and $\gamma$ to denote external legs from
each of the three branches of (Fig.\ref{3 point notation}), we find the translated
3-point vertex contribute to the expansion coefficient of (\ref{partial fraction3})
as 

\begin{equation}
\sum_{\alpha,\beta,\gamma}c\,\hat{\alpha}^{\frac{1}{2}}\hat{\beta}^{\frac{1}{2}}\eta_{\alpha}\eta_{\beta}+a\,\hat{\beta}^{\frac{1}{2}}\hat{\gamma}^{\frac{1}{2}}\eta_{\beta}\eta_{\gamma}+b\,\hat{\gamma}^{\frac{1}{2}}\hat{\alpha}^{\frac{1}{2}}\eta_{\gamma}\eta_{\alpha}\label{3 point partial expansion result}\end{equation}

where $a$, $b$ and $c$ here stand for
\begin{eqnarray}
&&
a=\widehat{l+1}+\cdots+\hat{n}+\hat{1} \\
&&
b=\hat{2}+\hat{3}+\cdots+\hat{k} \\
&&
c=\widehat{k+1}+\cdots+\hat{l}
\end{eqnarray}

The numerator of the formula (\ref{N1 Nair formula 2}) can accordingly
be written as

\begin{equation}
\sum_{\alpha,\beta,\gamma}\left\langle \alpha\beta\right\rangle \eta_{\alpha}\eta_{\beta}+\left\langle \beta\gamma\right\rangle \eta_{\beta}\eta_{\gamma}+\left\langle \gamma\delta\right\rangle \eta_{\gamma}\eta_{\alpha}\end{equation}

Applying (\ref{condition 3}) this becomes

\begin{equation}
\frac{\left(12\right)}{\hat{1}\hat{2}}\,\frac{1}{c}\,\left(\sum_{\alpha,\beta,\gamma}c\,\hat{\alpha}^{\frac{1}{2}}\hat{\beta}^{\frac{1}{2}}\eta_{\alpha}\eta_{\beta}+a\,\hat{\beta}^{\frac{1}{2}}\hat{\gamma}^{\frac{1}{2}}\eta_{\beta}\eta_{\gamma}+b\,\hat{\gamma}^{\frac{1}{2}}\hat{\alpha}^{\frac{1}{2}}\eta_{\gamma}\eta_{\alpha}\right)\end{equation}

The ratio between the coefficients of the $\eta_{\alpha}\eta_{\beta}$
term, $\eta_{\beta}\eta_{\gamma}$ term and $\eta_{\gamma}\eta_{\alpha}$
term are the same as the ratio in (\ref{3 point partial expansion result}).

\end{document}